\documentclass[pra,aps,amsmath,amssymb,amsfonts,twocolumn,nofootinbib,longbibliography]{revtex4}
\usepackage{}
\usepackage{amssymb}
\usepackage{bm,mathrsfs}
\usepackage{graphicx}
\usepackage{epsfig}
\usepackage{amsmath,bbm}
\usepackage{amsfonts,amssymb}
\usepackage{times}
\usepackage{verbatim}
\usepackage[sort&compress]{natbib}
\usepackage{amsmath}
\usepackage{bm}
\usepackage{float}
\usepackage{textgreek}
\allowdisplaybreaks[4]
\usepackage[colorlinks,breaklinks,linkcolor=blue,anchorcolor=blue,citecolor=blue,urlcolor=magenta,dvipdfm]{hyperref}

\usepackage{color}
\definecolor{zzz}{rgb}{0.9,0.0,0.4}
\begin{document}
\title{Parametric amplifier mediated nonreciprocal phonon laser in non-Markovian optomechanical resonators}
\author{H. Yi,$^{1}$ C. Cui,$^{1}$ J. Luo,$^{1}$ H. T. Cui,$^{2}$ Yan-Hui Zhou,$^{3,}$\footnote{\textcolor{zzz}{Contact author: yanhuizhou@126.com}} Cheng Shang,$^{4,}$ and H. Z. Shen$^{1,}$\footnote{\textcolor{zzz}{Contact author: shenhz458@nenu.edu.cn}}}

\affiliation{$^1$Center for Quantum Sciences and School of Physics, Northeast Normal University, Changchun 130024, China\\
$^2$School of Physics and Optoelectronic Engineering and Institute of Theoretical Physics, Ludong University, Yantai 264025, China\\
$^3$Quantum Information Research Center and Jiangxi Province Key Laboratory of Applied Optical Technology,
Shangrao Normal University, Shangrao 334001, China\\
$^4$Analytical quantum complexity RIKEN Hakubi Research Team,
RIKEN Center for Quantum Computing (RQC), Wako, Saitama 351-0198, Japan}

\date{\today}
\begin{abstract}
Phonon lasers amplify mechanical vibrations, serving as key tools for coherent acoustics research and applications such as ultrasensitive sensing and information processing, which has been extensively studied in Markovian systems but remains unexplored in non-Markovian ones. In this paper, we propose the nonreciprocal phonon laser in a coupled cavity system composed of nonlinear optomechanical resonator and degenerate optical parametric amplifier (DOPA) under the Markovian approximation. The mechanical gain and threshold power of the phonon laser are influenced by Sagnac-Fizeau effects, which leads to the unidirectional phonon laser and enhances the phonon laser performance via DOPA. Moreover, we generalize the nonreciprocal phonon laser to non-Markovian regimes. We find that non-Markovian effects can lead to the further enhancement of the phonon laser at a lower phonon laser threshold power with DOPA fixed. Our work reveals connections between low-power mechanical isolation and non-Markovian effects, offering a pathway to manipulate cavity optomechanical devices across a broader range of applications.
\end{abstract}

\maketitle
\section{Introduction}\label{Sec1}
Phonons possess unique advantages in certain aspects compared to photons, where phonons have a shorter wavelength at the same frequency. It enables higher resolution in applications such as imaging and precision machining. Since the first optical laser was invented in 1960 \cite{Maiman493}, researchers have been striving to accomplish phonon amplification by stimulated emission. In 2003, Chen and Khurgin proposed a feasible scheme for realizing a phonon laser \cite{Chen600}. Recently, a large amount of theoretical and experimental research related to phonon lasers has attracted the attention of researchers, including low threshold phonon lasers \cite{Ding2400350,Wang73,Ding15723,Wang083045}, PT-symmetric phonon lasers \cite{Xie22580,Wang43,Zhang479}, nonreciprocal phonon lasers \cite{Zhou2786,Lu260312,Chen052204,Wang282,Huang104201,Li044032,Wang41671}, polarization controlled phonon lasers \cite{Chafatinos4552,Papuccio17336}, and so on \cite{Jiang044205,Pan2400593,Mendonca33001,Meng115101,Maryam2184,Bhattacherjee639,Chen4144,Lozovik251,Lozovik431,
Makler316,Orlovskii316,Nam46,Yuan102201,Kemiktarak030802,Yuan012203,Xiao4234,Xiao28480,He37507,
He13998,Zhang14735,He031802,Xu053501,Ghamsari033511,Li013519,Liu044007,Lei054093,Camps125311,
Kabuss064305,Chudnovsky257205,Herrmann013004,Kabuss054301,Mizoguchi077402,Khaetskii186601,
Jing053604,Behrle043605,Wei110313,Pettit402,Huang372,Xiao17,Lozovik1307,Wang055108,Xiao245401,
Zadernovsky012005,Zhang093005,Mukhamedyanov20757,He9984,Yang12020,Dong193701,Karwat305, Eremeev2400497,Parsa046109,Kumar107135,Mercad073601,Zhang2207543119,Cui937,
Li16241,Djorwe1684,Karwat053855,Ransford043201,Droenner043805,Hammerer061803,Wu123022,
Xie053402,Mahboob127202,Okuyama013704,Rivlin322,Mukhamedyanov1822}. Soon after, phonon lasers have been experimentally realized in single trapped ions \cite{Vahala682}, semiconductor superlattices \cite{Kent215504}, and cavity optomechanical systems \cite{Grudinin083901}. Phonon lasers provide an important platform for quantum acoustics \cite{Ganesan033903,Chu199}, functional phononic devices \cite{Li1045,Shen010331,Wendt14385}, and nondestructive precision measurements or imaging \cite{Khurgin16,Swenson263511,Feng342}.

In mechanical engineering studies, the nonreciprocal transmission and amplification of phonons have attracted significant interest due to potential applications \cite{Wang194301}. Similar to optical nonreciprocity, nonreciprocal phononic devices based on nonreciprocal transmission and amplification of phonons have been employed to investigate chiral phonon transport or cooling \cite{Kim205,Xu65,Lai011502}, phonon isolation \cite{XuFleury516,Popa3398,Zhang115407,Devaux234301,Seif1207,Liang989}, one-way mechanical networks \cite{Shao267,Coulais461,Li035501,Wang4451,Penelet064012,Li084301,Poshakinskiy156801}, and backscattering-immune acoustic sensing or imaging \cite{Li962,Cummer495}. The nonreciprocal phonon laser by coupling an optomechanical resonator to a spinning optical resonator was proposed \cite{Jiang064037}, which gets nonreciprocity through the differences in Sagnac-Fizeau shifts when optical fields are input from different directions.

The combination of nonlinear optics and optomechanics provides a platform where amplified quantum effects can be generated \cite{Pina-Otey033835,Lin828}. Optical amplification can be obtained with introducing a DOPA pumped by an external laser into the optomechanical cavity \cite{Hu043824,Adamyan053818,L093602}. These features have enabled numerous applications, including enhanced entanglement \cite{Mi115,Hu24,Xuereb013809}, strong mechanical squeezing \cite{Agarwal043844}, enhanced optomechanical cooling \cite{Huang013821}, normal-mode splitting \cite{Huang033807}, control of photon blockade \cite{Shen013826,Shen023856,Shen043714}, and increased atom-cavity coupling \cite{Qin093601}. Moreover, the generation of down-converted photon pairs by DOPA exhibits nearly perfect single or dual squeezing. The directional quantum squeezing realizes an optical diode or circulator \cite{Tang083604}, paving the way for nonreciprocal devices that harness various quantum resources.  Based on one-way squeezing, nonreciprocal phonon lasers have also been explored \cite{Lu260312,Zhou2786}.

For open system \cite{Breuer2002,Weiss2012} interacting with environment, the Markovian approximation remains valid when the coupling between the system and environment is weak, where the characteristic times of the bath are sufficiently smaller than those of the quantum system. When the conditions are not met, it is necessary to investigate influences of non-Markovian effects on open quantum system properties \cite{Chang052105,Tan032102,Longhi063826,Leggett1,Breuer210401,Laine062115,Addis052103,Karlsson062108,
Xin053706,Wissmann042108,Shen053705,Lorenzo020102,Rivas050403,Shen023707,Luo044101,Wolf150402,Lu042103,Chruscinski120404,Shen012156,Zhang063853,Zhang083022,Xiong6053,Zhao29082,Triana183602,Cheng385,Shen062106,Cheng23678,Mu70,Li1363,Ding1814,
Sinha043603,Wu010601,Shen012107,Mu012334,Yang053712,Zhang033701,Li023712,Li5614,Cui032209,Shen032101,Shen31504,Shen2852,Cui042129,Shen052122,Shen033835,Diosi034101,Xiang053518,Zhao053518,Xiang045204}. Moreover, non-Markovian effects play meaningful roles in quantum information applications including quantum state engineering, quantum control, and quantum channel capacity \cite{Xue052304,Caruso1203,Arrigo211,Franco054304,Bylicka5720}, which can be realized in experiments \cite{Liu931,Trubarov7606,
Xiong032101,Cialdi052104,Khurana022107,Madsen233601,Guo230401,Li140501,Hoeppe043903,
Xu042328,Tang10002,Uriri052107,Anderson3202,Liu062208,Breuer021002,Vega015001}.

The above considerations lead us to investigate these issues: (i) Can nonreciprocal phonon lasers be manipulated by DOPA in spinning resonators? (ii) How can non-Markovian effects enhance nonreciprocal phonon lasers?

In this paper, we study the influences of DOPA on the nonreciprocal phonon laser in an optomechanical system under the Markovian approximation. We compare the differences in the nonreciprocal phonon laser when DOPA is driven by an optical field with different squeezing strengths. We show that the nonreciprocal phonon laser in spinning resonators can be enhanced via DOPA. Due to Sagnac-Fizeau shifts and DOPA, we find that the generation of the phonon laser can be controlled by adjusting the spinning speed of resonator and squeezing strength. Moreover, we extend the phonon laser scheme to non-Markovian regimes and compare the results with those under Markovian approximation. We show that the nonreciprocal phonon laser is further enhanced by non-Markovian effects, which is independent of the presence of DOPA. The scheme provides an effective way to control non-Markovian phonon lasers in a nonreciprocal optical device.

The rest of this paper is organized as follows. In Sec. \ref{Sec2},
we construct a system model with DOPA and derive the mechanical gain in Markovian environment. In Sec. \ref{Sec3}, we explore influences of DOPA and spinning speed on the nonreciprocal phonon laser. In Sec.~\ref{Sec4}, we extend the nonreciprocal phonon laser to non-Markovian regimes. In Sec.~\ref{Sec5}, we study the impacts of non-Markovian effects on the nonreciprocal phonon laser. The conclusions are summarized In Sec. \ref{Sec6}.
\section{Model and dynamics under Markovian approximation}\label{Sec2}
\begin{figure}[t]
\centerline{
\centering
\includegraphics[width=6.5cm, height=7cm, clip]{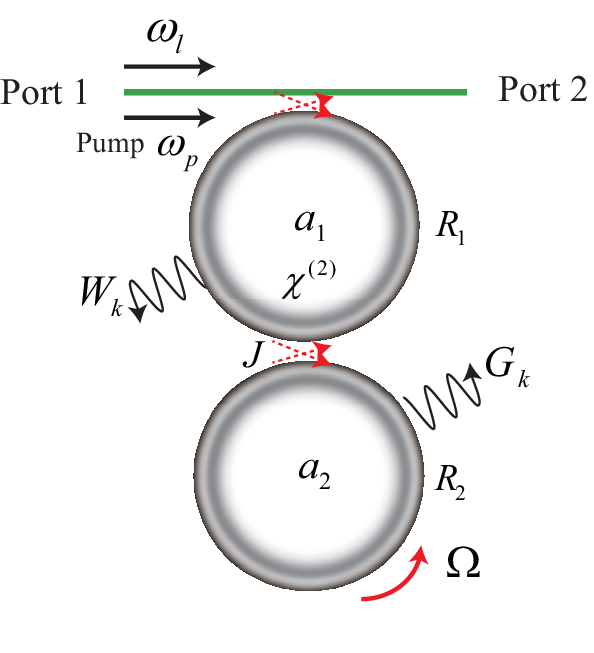}}
\caption{The figure shows a nonreciprocal phonon laser device. $R_1$ is a nonlinear optical resonator, which supports a mechanical breathing mode. The strong driving field with frequency $\omega_l$ and pump field (excited by $\chi^{(2)}$ nonlinearity) with frequency $\omega_p$ are input from port 1 exciting the clockwise mode in $R_1$. $R_2$ is an optical resonator with angular velocity $\Omega$. The resonators $R_1$ and $R_2$ interact with the strength $J$. The two photon modes couple to two non-Markovian environments with eigenfrequencies $x_{k}$ and $n_k$ (annihilation operators $\hat{d}_k$ and $\hat{e}_k$), where ${W}_k$ and ${G}_k$ denote coupling coefficients between two resonators and two environments. Sec. \ref{Sec2} and Sec. \ref{Sec3} consider the Markovian case, while Sec. \ref{Sec4} and Sec. \ref{Sec5} focus on the non-Markovian one.}\label{model}
\end{figure}

The system in Fig.~\ref{model} comprises two coupled whispering gallery-mode (WGM) microtoroid resonators, in which $R_1$ (with frequency $\omega_1$) supports a mechanical breathing mode (with frequency $\omega_b$ and effective mass $m$) driven by light with frequency $\omega_l$ and amplitude $\varepsilon$. We introduce a pump field to induce the squeezed mode $a_s$ by driving a $\chi^{(2)}$ nonlinear medium from port 1, which can enhance the phonon laser. Moreover, while enhancing the phonon laser, the classical coherent pump input introduces effective thermalization noise to the squeezed mode $a_s$. Since phonon lasers are realized in the classical regime with a strong driving field, the effective thermalization noise of the squeezed mode $a_s$ can be neglected in the following discussion and squeezed mode $a_s$ has the same decay rate with the bare mode $a_1$. The another resonator $R_2$ is an optical resonator, which supports a mode $a_2$ with frequency $\omega_2$ and decay rate $\Gamma_2$.

It is well known that a resonator supports two optical modes called CW and CCW. When the resonator is spinning, the frequencies of CW and CCW modes experience Sagnac-Fizeau shifts, which lead to nonreciprocity. For light propagating through different ports, we have
\begin{equation}
\begin{aligned}
{\Delta _{{\rm{sag}}}} =  \pm \Omega \frac{{n{r_2}{\omega _2}}}{c}\left( {1 - \frac{1}{{{n^2}}} - \frac{\lambda }{n}\frac{{dn}}{{d\lambda }}} \right),
\end{aligned}
\end{equation}
where $\Omega$ represents the angular velocity of the spinning resonator. $\lambda  = c/{\omega _2}$ is the optical wavelength. $c$ is the speed of light in vacuum. $n$ and $r_2$ denote the refractive index and radius of the resonator $R_2$, respectively. Since the dispersion term $\left( {\lambda /n} \right)\left( {dn/d\lambda } \right)$-the Lorentz correction to the Fresnel-Fizeau drag coefficient-is small, it can generally be neglected for most common materials \cite{Maayani558569}. When the resonator $R_1$ is driven from port 1, the resonator spins in the same direction as the optical wave rotation, which denotes $\Delta_{\rm{sag}} > 0$. In contrast, $\Delta_{\rm{sag}} < 0$ represents the case where the resonator is driven from port 2, which corresponds to the resonator spinning in the opposite direction as the optical wave rotation. Below we show that for our optomechanical system the mechanical gain exhibits distinct changes due to DOPA and spinning of the resonator, hence resulting in an enhanced nonreciprocal phonon laser. The system Hamiltonian reads
\begin{equation}
\begin{aligned}
{{\hat H_S}} =& {\omega _1}{{\hat a_1}^\dag }\hat a_1 + {\omega _2}{{\hat a_2}^\dag }\hat a_2+ \omega _b{{\hat b}^\dag }\hat b + J({\hat a_1}^\dag {\hat a_2}+ {\hat a_2}^\dag {\hat a_1})\\
&- g x_0 {{\hat a_1}^\dag }\hat a_1({\hat b}^\dag+ \hat b)+ i \varepsilon({\hat a_1}^\dag e^{-i\phi}e^{-i \omega_l t}-{\hat a_1}e^{i\phi}e^{i \omega_l t})\\
&+\frac{\Lambda}{2}({\hat a_1}^{\dag 2} e^{-i\theta-i\omega_pt}+{\hat a_1}^{ 2 } e^{i\theta+i\omega_pt}),\label{H0}
\end{aligned}
\end{equation}
where $\hat a_1$ ($\hat a_1^\dag$) and $\hat a_2$ ($\hat a_2^\dag$) denote the annihilation (creation) operators of photon modes in $R_1$ and $R_2$, respectively. The $\hat b$ ($\hat b^\dag$) is the annihilation (creation) operator of the phonon mode in $R_1$. The first and second terms of the Hamiltonian in Eq.~(\ref{H0}) respectively describe different photon modes in spinning resonators, while the third term describes the phonon mode. The fourth (fifth) term denotes the interaction between photon mode $ a_2$ (phonon mode $ b$) and photon mode $ a_1$ with the coupling strength $J$ ($g=\omega_1/r_1$) and ${x_0} = \sqrt {\hbar /2m{\omega _b}} $. The driving field with the amplitude ${\varepsilon} = \sqrt {{\Gamma _1}{P_{in}}/\hbar {\omega _l}} $ (frequency $\omega_l$) and the phase $\phi$ is described by the sixth term in Eq.~(\ref{H0}). $\Gamma_1$ is the decay rate of the photon mode $ a_1$, while $P_{in}$ is the input power of the driving field. The last term describes the pump field with squeezing strength $\Lambda$, phase $\theta$, and pump frequency $\omega_p$. The pump phase $\theta$ in Eq.~(\ref{H0}) can be eliminated via the unitary transformation $a_1 \to a_1$ $e^{-i\theta/2}$ and $a_2 \to a_2$ $e^{-i\theta/2}$ by setting $\theta=2\phi$. Therefore, without loss of generality, we set $\theta=0$ in the following. The Hamiltonian in Eq.~(\ref{H0}) with the frame rotating becomes
\begin{equation}
\begin{aligned}
{{\hat H_{1}}} =& {-\Delta_e}{{\hat a_1}^\dag }\hat a_1 - ({\Delta_l+\Delta_{\rm{sag}}}){{\hat a_2}^\dag }\hat a_2+ \omega _b{{\hat b}^\dag }\hat b \\
&+ J({\hat a_1}^\dag {\hat a_2}+ {\hat a_2}^\dag {\hat a_1})+ i \varepsilon({\hat a_1}^\dag e^{-i \Delta_1 t}-{\hat a_1} e^{i \Delta_1 t})\\
&- g x_0 {{\hat a_1}^\dag }\hat a_1({\hat b}^\dag+ \hat b) +\frac{\Lambda}{2}({\hat a_1}^{\dag 2} +{\hat a_1}^{ 2 }),\label{Heff1}
\end{aligned}
\end{equation}
where ${\Delta _e} = {\omega _e} - {\omega _1}$, ${\Delta _l} = {\omega _e} - {\omega _2}$, ${\Delta _1} = {\omega _l} - {\omega _e}$, and $\omega_e=\omega_p/2$. We use the squeezed operator $\hat a_s$ via the Bogoliubov transformation to diagonalize the Hamiltonian in Eq.~(\ref{Heff1})
\begin{equation}
\begin{aligned}
\hat a_s=&\cosh (r)\hat a_1+ \sinh(r)\hat a_1^\dag,\label{as}
\end{aligned}
\end{equation}
where the squeezing parameter $r=\ln [ (\Delta _e - \Lambda)/(\Delta _e+ \Lambda) ]/4$. Substituting Eq.~(\ref{as}) into Eq.~(\ref{Heff1}) causes
\begin{equation}
\begin{aligned}
\hat H_2 =&  - ({\Delta _l+\Delta_{\rm{sag}}})\hat a_2^\dag {\hat a_2}- g{x_0}{\sinh ^2}(r)(\hat b + {\hat b^\dag })\\
 &+ {\omega _b}{\hat b^\dag }\hat b - g{x_0}\cosh (2r)\hat a_s^\dag {\hat a_s}(\hat b + {\hat b^\dag })\\
 &+ g{x_0}\sinh (2r)(\hat a_s^{\dag 2}+ \hat a_s^2)(\hat b + {\hat b^\dag })/2 \\
 &- \{{\Delta _e}[{\cosh ^2}(r) + {\sinh ^2}(r)]\\
 &+ 2\Lambda \cosh (r)\sinh (r)\}\hat a_s^\dag {\hat a_s}\\
 &- {\Delta _e}{\sinh ^2}(r) - \Lambda \cosh (r)\sinh (r)\\
  &+ i\varepsilon \{[\hat a_s^\dag \cosh (r) - {\hat a_s}\sinh (r)] e^{-i \Delta_1 t} \\
  &- [{\hat a_s \cosh (r) - {\hat a_s^\dag}\sinh (r)}]e^{i \Delta_1 t}\}\\
  &+ J\cosh (r)(\hat a_2^\dag {\hat a_s} + \hat a_s^\dag {\hat a_2})\\
 &- J\sinh (r)(\hat a_2^\dag \hat a_s^\dag  + {\hat a_2}{\hat a_s}).\label{H2}
\end{aligned}
\end{equation}
\begin{figure}[t]
\centerline{
\centering
\includegraphics[width=8cm, height=6cm, clip]{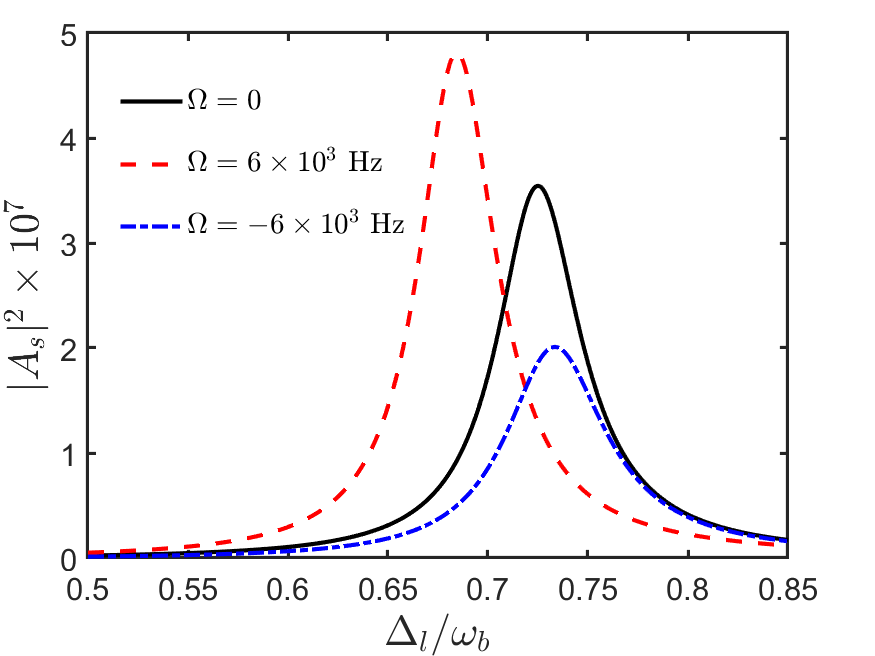}}
\caption{The steady-state photon number in Eq.~(\ref{eqas}) is plotted as a
function of the detuning $\Delta_l$. The red dashed and blue dash-dotted lines respectively correspond to $\Omega=6 \times 10^3$ Hz and $\Omega=-6 \times 10^3$ Hz. The black solid line denotes $\Omega=0$, which implies that the resonant is stationary. The parameters chosen are $n=1.48$, $\omega_1/{2 \pi}=\omega_2/{2 \pi}=30.7$ THz, $r_1=33$ $\mu$m, $Q_1=9.7\times10^7$, $r_2=4.75$ mm, $Q_2=3\times10^7$, $m=50$ ng, $\gamma_b=0.24$ MHz, $J=0.48\omega_b$, $\omega_b / 2\pi=23.4$ MHz, $\Delta_1=-0.2\omega_b$, $P_{in}=10$ $\mu$W, and $\Lambda=0.3\omega_b$.}\label{Fig2}
\end{figure}With the rotating wave approximation, we neglect terms $g{x_0}\sinh (2r)(\hat a_s^{\dag 2}+ \hat a_s^2)(\hat b + {\hat b^\dag })/2, g{x_0}{\sinh ^2}(r)(\hat b + {\hat b^\dag }), J\sinh (r)(\hat a_2^\dag \hat a_s^\dag  + {\hat a_2}{\hat a_s}), i\varepsilon{\hat a_s^\dag}\sinh (r) e^{i \Delta_1 t}$, and $i\varepsilon{\hat a_s}\sinh (r) e^{-i \Delta_1 t}$ \cite{L093602,Lu260312}. The effective Hamiltonian after squeezed transformation from Eq.~(\ref{H2}) for the forward-input case is expressed as
\begin{equation}
\begin{aligned}
{{\hat H_{F}}} =& -{\Delta _s}{{\hat a_s}^\dag }\hat a_s - ({\Delta _l+\Delta_{\rm{sag}}}){{\hat a_2}^\dag }\hat a_2+ \omega _b{{\hat b}^\dag }\hat b\\
&+ J_s({\hat a_s}^\dag {\hat a_2}+ {\hat a_2}^\dag {\hat a_s})- g_s x_0 {{\hat a_s}^\dag }\hat a_s({\hat b}^\dag+ \hat b)\\&+ i \varepsilon_s({\hat a_s}^\dag e^{-i \Delta_1 t}-{\hat a_s} e^{i \Delta_1 t}),\label{Heff}
\end{aligned}
\end{equation}
where $\Delta_s=(\Delta_e+\Lambda)e^{2r}$, $J_s=J \cosh{(r)}$, $g_s=g \cosh(2r)$, and $\varepsilon_s=\varepsilon \cosh{(r)}$. In the backward-input case, the optical mode in $R_1$ is not squeezed, which corresponds to squeezing strength of zero. In strong driving conditions, we are only interested in the mean-number behaviors and define $a_s \equiv \langle \hat{a}_s \rangle$, $a_2 \equiv \langle \hat{a}_2 \rangle$, and $ b \equiv \langle \hat{b} \rangle$. Eq.~(\ref{Heff}) gives
\begin{equation}
\begin{aligned}
\dot{a}_s=&(i\Delta_s-\Gamma_1/2){a_s}+i g_s x_0 a_s({b+ {b}^*})\\
&-i J_s a_2+\varepsilon_s e^{-i \Delta_1 t},\\
\dot{a}_2=&(i\Delta_l-\Gamma_2/2+i\Delta_{\rm{sag}}){a_2}-i J_s {a_s},\\
\dot{b}=&-(i\omega_b+\gamma_b)b+ i g_s x_0 {{a_s^*} }a_s,\label{eqa}
\end{aligned}
\end{equation}
where $\Gamma_1=2\omega_1/Q_1$ and $\Gamma_2=2\omega_2/Q_2$. $\gamma_b$ is the damping rate of the mechanical mode $b$. Introducing ${a_ s } = {A_s}{e^{ - i{\Delta _1}t}}$ and ${a_ 2 } = {A_2}{e^{ - i{\Delta _1}t}}$, Eq.~(\ref{eqa}) becomes

\begin{equation}
\begin{aligned}
{\ddot { A}_s}   =& 2i{\Delta _1}{\dot { A}_s}+{\Delta _1^2}{ A_s}+(i{\Delta _s} - {\Gamma _1/2})({\dot { A}_s} - i{\Delta _1}{ A_s})\\
&+ i{g_s}{x_s}({\dot { A}_s}- i{\Delta _1}{ A_s})+ i{g_s}\dot{x}_s A_s\\
&- i{\Delta _1}{\varepsilon _s} - i{J_s}({\dot { A}_2} - i{\Delta _1}{ A_2}),\\
{\ddot A_2}  =&  2i{\Delta _1}{\dot A_2}+ {\Delta _1^2}{A_2}- i{J_s}({\dot A_s} - i{\Delta _1}{A_s})\\
&+(i{\Delta _l}+ i{\Delta _{\rm{sag}}}- {\Gamma_2/2})({\dot A_2} - i{\Delta _1}{A_2}),
\end{aligned}
\end{equation}
which leads to the steady-state solutions as
\begin{equation}
\begin{aligned}
A_s=&\frac{i \varepsilon_s (i \Gamma_2/2+\Delta_A)}{-J_s^2+(i \Gamma_2/2+\Delta_A)(i \Gamma_1/2+\Delta_s+ x_s g_s+\Delta_1)},\\
A_2=&\frac{i J_s \varepsilon_s  }{-J_s^2+(i \Gamma_2/2+\Delta_A)(i \Gamma_1/2+\Delta_s+ x_s g_s+\Delta_1)},\\
b=&\frac{ x_0 g_s |A_s|^2}{-i \gamma_b+ \omega_b},\label{eqas}\\
\end{aligned}
\end{equation}
where $\Delta_A=\Delta_l +\Delta_1+\Delta_{\rm{sag}}$. The steady-state mechanical displacement $x_s$ is given by
\begin{equation}
\begin{aligned}
{x_s}=x_0(b+b^*) = \frac{{\hbar g_s{{\left| {{A_s}} \right|}^2}}}{{m\left( {\omega _b^2 + \gamma _b^2} \right)}},
\end{aligned}
\end{equation}
which is related to the steady-state photon number $|A_s|^2$, depending on the $\Delta_{\rm{sag}}$. The steady-state population of the photon mode is investigated as a function of the detuning $\Delta_l$ in Fig.~\ref{Fig2}. We find that light propagating through different ports results in different steady-state photon numbers $|A_s|^2$, which increase for $\Delta_{\rm{sag}}>0$ and decrease for $\Delta_{\rm{sag}}<0$. The influence of the intracavity photon population is to change the radiation pressure, which in turn modifies the optomechanical coupling strength. Different steady-state photon numbers lead to different mechanical displacements, which is the origin of the nonreciprocal phonon laser. Moreover, the ratio $\eta$ is the mechanical displacement amplification factor
\begin{equation}
\begin{aligned}
\eta  = \frac{{{x_s}(\Omega  \ne 0)}}{{{x_s}(\Omega  = 0)}}\label{eta},
\end{aligned}
\end{equation}
which is defined to describe the enhancement in steady-state mechanical displacement. In Fig.~\ref{Fig3}, the mechanical displacement amplification factor $\eta$ in Eq.~(\ref{eta}) is plotted as a function of the detuning $\Delta_l$ and angular velocity $\Omega$. Due to Sagnac-Fizeau shifts and DOPA, the mechanical displacement varies with the direction of the input driving field. Specifically, when the driving field is input from port 1, the mechanical displacement amplification factor peaks at $\Delta_l=0.67\omega_b$, as indicated by the red dashed line in Fig.~\ref{Fig3}. On the contrary, the mechanical displacement is suppressed by the input of optical field from port 2 at the same detuning. Similar to the optical laser, we introduce the supermode operators $\hat a_+=(\hat a_s+\hat a_2)/\sqrt{2}$ and $\hat a_-=(\hat a_s-\hat a_2)/\sqrt{2}$ satisfying $[\hat a_+,\hat a_+^\dag]=[\hat a_-,\hat a_-^\dag]=1$. With this, the Hamiltonian in Eq.~(\ref{Heff}) becomes (see Appendix \ref{DSH} for the detailed derivation)
\begin{figure}[t]
\centerline{
\centering
\includegraphics[width=8cm, height=6cm, clip]{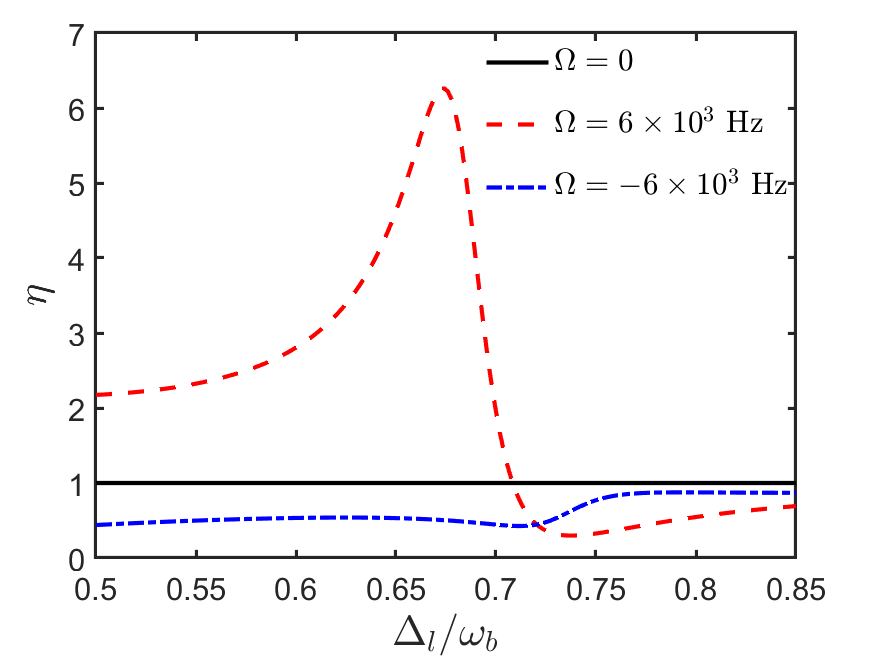}}
\caption{The mechanical displacement amplification factor $\eta$ in Eq.~(\ref{eta}) is plotted as a function of the detuning $\Delta_l$. It exhibits a dependence of steady-state mechanical displacement $x_s$ on the angular velocity $\Omega$. The parameters chosen are same as Fig.~\ref{Fig2}.}\label{Fig3}
\end{figure}
\begin{equation}
\begin{aligned}
{{\hat H_{s}}} =& {\omega _+}{{\hat a_+}^\dag }\hat a_+ + {\omega _-}{{\hat a_-}^\dag }\hat a_-+ \frac{\Delta+\Delta_{\rm{sag}}}{2}({\hat a_+}^\dag {\hat a_-}+ {\hat a_-}^\dag {\hat a_+})\\
&+ \omega _b{{\hat b}^\dag }\hat b- \frac{g_s x_0}{2} ({{\hat a_+}^\dag }\hat a_- \hat b+{\hat b}^\dag{\hat a_-}^\dag \hat a_+)\\
&+ \frac{i}{\sqrt{2}} \varepsilon_s[({\hat a_+}^\dag +{\hat a_-}^\dag)e^{-i \Delta_1 t}-({\hat a_+} +{\hat a_-}) e^{i \Delta_1 t})].\label{He+}
\end{aligned}
\end{equation}
In the supermode picture, with $a_+ \equiv  \langle \hat{a}_+ \rangle$, $a_- \equiv  \langle \hat{a}_- \rangle$, and $b \equiv  \langle \hat{b} \rangle$, we obtain from Eq.~(\ref{He+})
\begin{equation}
\begin{aligned}
\dot{a}_+=&-(i \omega_+ + \Gamma_0/2)a_+ +\frac{i}{2}(g_s x_0 b-\Delta\\
&-\Delta_{\rm{sag}})a_-+\frac{\varepsilon_s}{\sqrt{2}}e^{-i \Delta_1 t},\\
\dot{a}_-=&-(i \omega_- + \Gamma_0/2)a_- +\frac{i}{2}(g_s x_0 b^*-\Delta\\
&-\Delta_{\rm{sag}})a_++\frac{\varepsilon_s}{\sqrt{2}}e^{-i \Delta_1 t},\label{A1}
\end{aligned}
\end{equation}
where $\omega_+=-(\Delta_s+\Delta_l)/{2}-\Delta_{\rm{sag}}/{2}+J_s$,
$\omega_-=-(\Delta_s+\Delta_l)/{2}-\Delta_{\rm{sag}}/{2}-J_s$, $\Gamma_0=(\Gamma_1+\Gamma_2)/2$, and $\Delta=\Delta_l-\Delta_s$.
According to Eq.~(\ref{A1}), the steady-state solutions of ${A_ + } = {a_+}{e^{ i{\Delta _1}t}}$ and ${A_- } = {a_-}{e^{  i{\Delta _1}t}}$ are given by
\begin{equation}
\begin{aligned}
A_+=&\frac{i \varepsilon_s (-i\Gamma_0-\Delta-2\Delta_1-\Delta_{\rm{sag}}+b x_0 g_s+2\omega_-)}{\sqrt{2}[2\beta-i\Gamma_0(2 \Delta_1+\Delta_l + \Delta_s + \Delta_{\rm{sag}})]},\\
A_-=&\frac{i \varepsilon_s (-i\Gamma_0-\Delta-2\Delta_1-\Delta_{\rm{sag}}+b^* x_0 g_s+2\omega_+)}{\sqrt{2}[2\beta-i\Gamma_0(2 \Delta_1+\Delta_l + \Delta_s + \Delta_{\rm{sag}})]},\label{a+1}
\end{aligned}
\end{equation}
where
\begin{equation}
\begin{aligned}
\beta=&J_s^2+\frac{\Gamma_0^2}{4}-\Delta_1^2-\Delta_s \Delta_l+\frac{x_0^2 g_s^2 n_b}{4}\\
&+\frac{1}{2}{\mathop{\rm Re}\nolimits}(b) x_0 g_s(\Delta_s-\Delta_l-\Delta_{\rm{sag}})\\
&-\Delta_1(\Delta_l+\Delta_s+\Delta_{\rm{sag}})-\Delta_s \Delta_{\rm{sag}},
\end{aligned}
\end{equation}
with $n_b=\langle \hat b^\dag \hat b \rangle$ and Re($b$) respectively denoting the phonon number and real part of $b$. According to definitions $ p=\langle \hat a_-^\dag \hat a_+ \rangle$ and $\delta n=\langle \hat a_+^\dag \hat a_+ \rangle -\langle \hat a_-^\dag \hat a_-\rangle$ (population difference between two supermodes), Eq.~(\ref{He+}) triggers
\begin{equation}
\begin{aligned}
\dot{b}=&-(i \omega_b+\gamma_b)b+\frac{i}{2}g_s x_0 p,\\
\dot{p}=&-2(\frac{\Gamma_0}{2}+i J_s)p+\frac{i}{2}(\Delta+\Delta_{\rm{sag}}-g_s x_0 b)\delta{n}\\&+\frac{1}{\sqrt{2}}\varepsilon_s(A_+ + {A_-^*}).\label{p1}
\end{aligned}
\end{equation}
According to Eq.~(\ref{p1}), we obtain the steady-state value of $p$ with $\Gamma_1, \Gamma_2 \gg \gamma_b$
\begin{equation}
\begin{aligned}
p=\frac{-i\sqrt{2}(A_-^*+A_+) \varepsilon_s+\delta{n}(\Delta_{\rm{sag}}+\Delta-b x_0 g_s)}{4J_s-2i \Gamma_0-2 \omega_b}.\label{p3}
\end{aligned}
\end{equation}
\begin{figure}[t]
\centerline{
\centering
\includegraphics[width=8cm, height=6cm, clip]{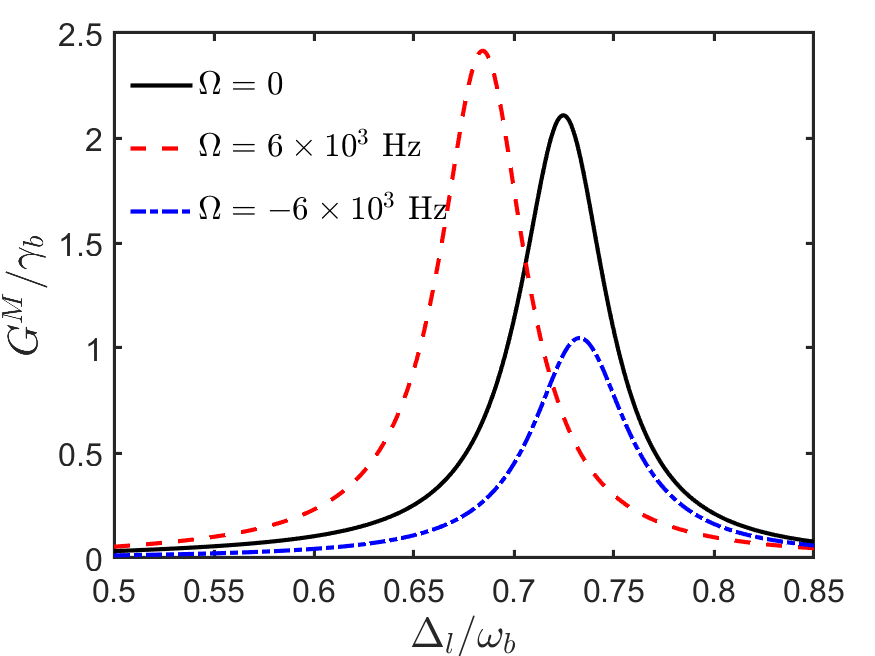}}
\caption{The mechanical gain $G^M$ in Eq.~({\ref{G}}) is plotted as a function of the detuning $\Delta_l$. The phonon laser can be realized when $G^M/\gamma_b>1$. The parameters chosen are same as Fig.~\ref{Fig2}.}\label{Fig4}
\end{figure}Substituting Eqs.~(\ref{a+1}) and (\ref{p3}) into Eq.~(\ref{p1}) gets
\begin{equation}
\begin{aligned}
\dot b = ( - i{\omega _b} - i{\omega ^M} + G^M - {\gamma _b})b + F^M,\label{b3}
\end{aligned}
\end{equation}
where the mechanical gain
\begin{equation}
\begin{aligned}
G^M=&G^M_0+G^M_1,\\
G^M_0=&\frac{x_0^2 \Gamma_0 \delta n g_s^2}{4\Gamma_0^2+(4 J_s-2\omega_b)^2},\\
G^M_1=&\frac{x_0^2 \Gamma_0\Delta_g\varepsilon_s^2 g_s^2(\omega_b-2J_s)}{2(4\beta^2+\Gamma_0^2\Delta_g^2)[\Gamma_0^2+(-2J_s+\omega_b)^2]},\label{G}
\end{aligned}
\end{equation}
where $\Delta_g=2\Delta_1+\Delta_l+\Delta_s+\Delta_{\rm{sag}}$. The mechanical gain $G^M$ given by Eq.~(\ref{G}) consists of two components: $G^M_0$ and $G^M_1$. $G^M_0$ is proportional to the population inversion $\delta n$, while $G^M_1$ arises from the drive terms for the unpaired supermodes in Eq.~(\ref{A1}). When $G^M > \gamma_b$, the phonon mode can be amplified. Note that both $G^M_0$ and $G^M_1$ depend on $\Delta_{\rm{sag}}$ and squeezing strength $\Lambda$. It means that a nonreciprocal phonon laser can be obtained by tuning $\Delta_{\rm{sag}}$ and $\Lambda$. Given the lack of connections to the phonon laser threshold power, expressions for $\omega ^M$ and $F^M$ in Eq.~(\ref{b3}) are not listed for convenience.


In Fig.~\ref{Fig4}, we demonstrate the directional dependence of the mechanical gain $G^M$. The black solid line shows the result for a stationary optical resonator ($\Omega=0$), where the mechanical gain is independent of the propagation direction of the driving light. In contrast, we find the peak position of the mechanical gain has different changes depending on whether the driving fields are input from different ports when the resonator is spinning. That is, the mechanical gain is enhanced in one direction but suppressed in another direction in Fig.~\ref{Fig4}. For $\Omega=6 \times10^3$ Hz, the peak value of the mechanical gain is $2.42\gamma_b$, while the mechanical gain is only $0.27\omega_b$ for driving from the opposite direction at the same frequency. This nonreciprocity can be attributed to the fact that the spinning of the resonator affects the strength of the optomechanical interaction, leading to different mechanical gains. By adjusting both the speed and direction of spinning, the mechanical gain $G^M$ can be effectively tuned, enabling the realization of the phonon laser across different operational regimes. With the threshold condition $G^M = \gamma_b$ for the phonon laser, the stimulated emitted phonon number is defined as
\begin{figure}[t]
\centerline{
\centering
\includegraphics[width=8cm, height=6cm, clip]{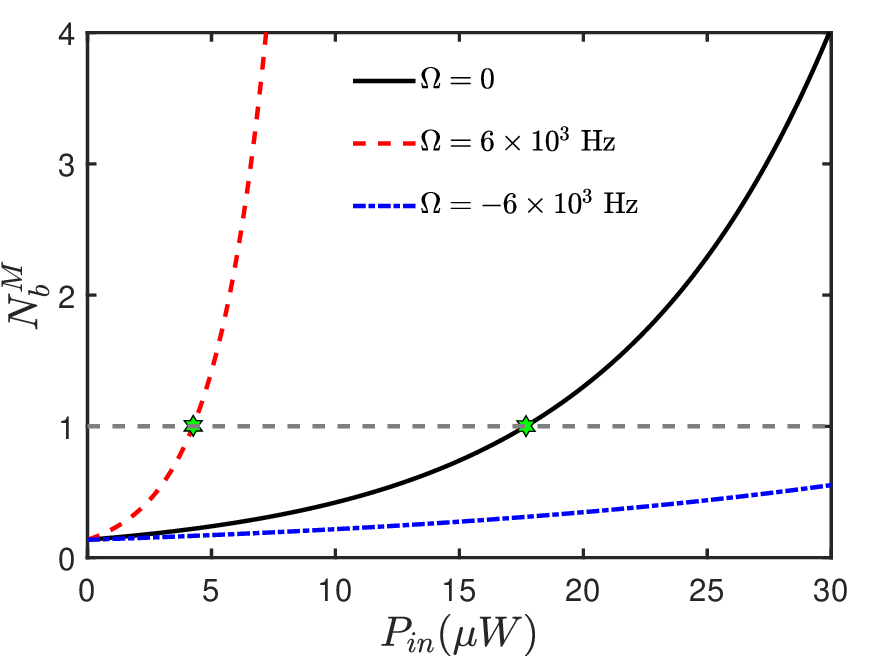}}
\caption{The stimulated emitted phonon number $N^M_b$ in Eq.~(\ref{Nb}) is plotted as a function of the pump power $P_{in}$. The six-pointed stars mark the threshold power, which is determined by using the threshold condition $G^M=\gamma_b$ in Eq.~(\ref{Nb}). The parameter chosen is $\Delta_l=0.68\omega_b$, while other parameters are same as Fig.~\ref{Fig2}.}\label{Fig5}
\end{figure}
\begin{equation}
\begin{aligned}
{N^M_b} = \exp [2(G^M - \gamma_b)/\gamma_b],\label{Nb}
\end{aligned}
\end{equation}
which characterizes the performance of the phonon laser. By setting the stimulated emitted phonon number $N^M_b=1$, we can obtain the threshold pump power.

Figure~\ref{Fig5} shows the relationship between the phonon number under stimulated emission $N^M_b$ and pump power $P_{in}$. We find that the threshold power is strongly influenced by the spinning speed and direction of the resonator. Compared to the case without the spinning of the resonator, the threshold pump power can be reduced from 17.7 $\mu$W to 4.26 $\mu$W for $\Omega=6 \times 10^3$ Hz due to the enhancement of the mechanical gain. This implies that the phonon laser can be generated at a lower power level. However, for $\Omega=-6 \times 10^3$ Hz, phonon amplification can not be realized because of the reduction in mechanical gain, where the phonon laser requires a larger input power to obtain phonon amplification.
\section{DISCUSSIONS on DOPA-enhanced nonreciprocal phonon laser}\label{Sec3}
\begin{figure}[h]
\centerline{
\centering
\includegraphics[width=8cm, height=6cm, clip]{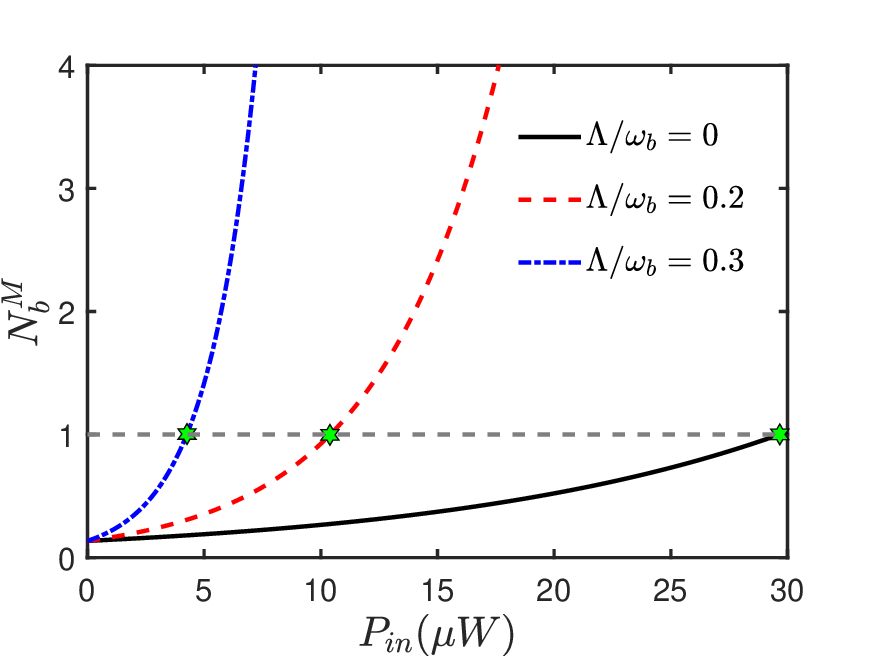}}
\caption{The stimulated emitted phonon number $N^M_b$ in Eq.~(\ref{Nb}) is plotted as a function of the pump power $P_{in}$ for different squeezing strengths. The six-pointed stars mark the threshold power, which is determined by using the threshold condition $G^M=\gamma_b$ in Eq.~(\ref{Nb}). The parameters chosen are $\Omega=6\times10^3$ Hz and $\Delta_l=0.68\omega_b$, while other parameters are same as Fig.~\ref{Fig2}.}\label{Fig6}
\end{figure}
In numerical simulations, we select experimentally feasible parameters, which are $n=1.48$, $\omega_1/{2 \pi}=\omega_2/{2 \pi}=30.7$ THz, $r_1=33$ $\mu$m, $Q_1=9.7\times10^7$, $r_2=4.75$ mm, $Q_2=3\times10^7$, $m=50$ ng, $\gamma_b=0.24$ MHz, $J=0.48\omega_b$, $\omega_b / 2\pi=23.4$ MHz, $\Delta_1=-0.2\omega_b$, $P_{in}=10$ $\mu$W, and $\Lambda=0.3\omega_b$ \cite{Jiang064037,Grudinin083901}. In our system, two distinct mechanisms contribute the nonreciprocity. The first is the directional squeezing induced by the DOPA, which is pumped only from port 1. This unidirectional squeezing enhances the effective parameters (\(J_s\), \(g_s\), \(\epsilon_s\)) for the forward direction while leaving the backward direction unchanged, thus producing the nonreciprocal gain. The second mechanism is the Sagnac-Fizeau effect from the spinning resonator, which shifts the optical resonance frequencies by \(\pm\Delta_{\mathrm{sag}}\) for the two propagation directions. The interplay of these two effects enables the strong nonreciprocal phonon laser performance discussed below. To study the influences of DOPA on phonon amplification, the stimulated emitted phonon number $N^M_b$ in Eq.~(\ref{Nb}) is plotted as a function of the pump power $P_{in}$ for different squeezing strengths $\Lambda$ in Fig.~\ref{Fig6}. In the absence of a pump field driving DOPA, as represented by the black solid line in Fig.~\ref{Fig6}, the threshold pump power of the phonon laser is 29.7 $\mu$W. As the squeezing strength increases, the pump threshold power decreases continuously. At the squeezing strength $\Lambda=0.3\omega_b$, the pump threshold power arrives at 4.26 $\mu$W. The underlying physical mechanism can be explained as the squeezing effect inducing an enhancement in coupling strengths, which leads to the mechanical gain $G^M$ in Eq.~({\ref{G}}) exceeding damping rate $\gamma_b$. This indicates that the phonon laser can be tuned by the spinning of the optical resonator and squeezing strength. To investigate the influences between these two control parameters, we introduce two isolation parameters
\begin{figure}[h]
\centerline{
\centering
\includegraphics[width=8cm, height=6cm, clip]{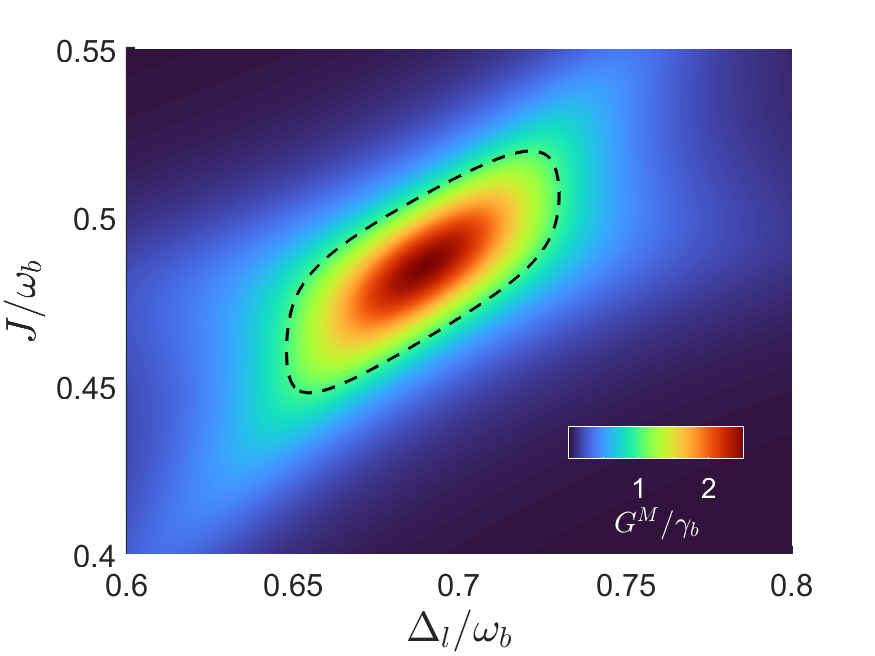}}
\caption{The mechanical gain $G^M$ in Eq.~(\ref{G}) is plotted as a function of the detuning $\Delta_l$ and coupling strength $J$. The black dashed line corresponds to $G^M/\gamma_b=1$, which is the threshold condition for generating a phonon laser. The parameter chosen is $\Omega=6\times10^3$ Hz, while other parameters chosen are same as Fig.~\ref{Fig2}.}\label{Fig7}
\end{figure}
\begin{equation}
\begin{aligned}
R_\Omega &= 10~{\log _{10}}\frac{{{N^M_b}(\Omega  \ne 0)}}{{{N^M_b}(\Omega  = 0)}}, \,\,\,\,R_\Lambda = 10~{\log _{10}}\frac{{{N^M_b}(\Lambda  \ne 0)}}{{{N^M_b}(\Lambda  = 0)}}.\label{R}
\end{aligned}
\end{equation}
\begin{figure}[t]
\centerline{
\centering
\includegraphics[width=8cm, height=6cm, clip]{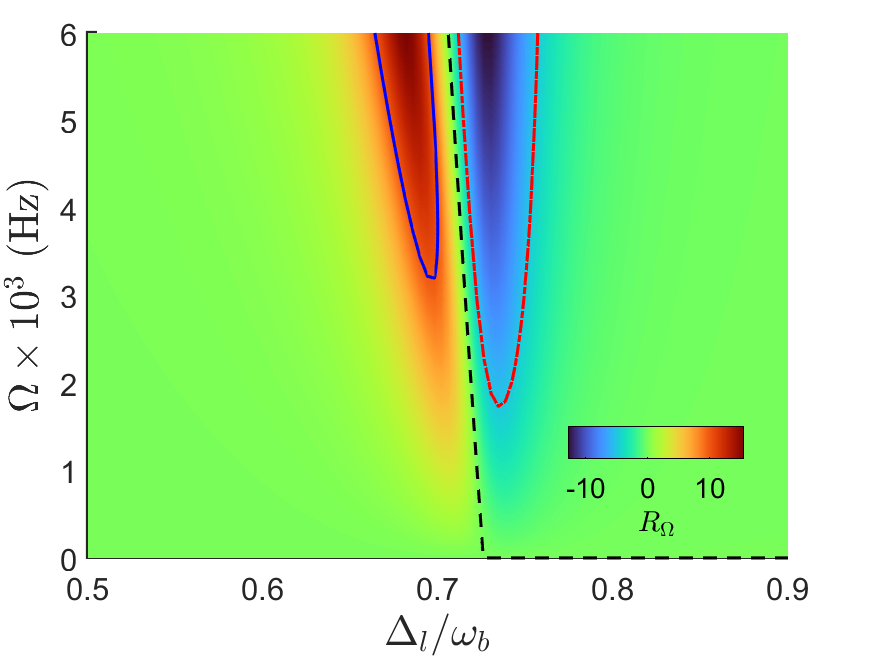}}
\caption{The isolation parameter $R_\Omega$ in Eq.~(\ref{R}) is plotted as a function of the detuning $\Delta_l$ and angular velocity $\Omega$. The black dashed, red dash-dotted, and blue solid lines respectively correspond to $R_\Omega=0$, $R_\Omega=-5$, and $R_\Omega=10$. The parameter chosen is $\Lambda=0.3\omega_b$, while other parameters chosen are same as Fig.~\ref{Fig2}.}\label{Fig8}
\end{figure}

In Fig.~\ref{Fig7}, we plot the mechanical gain $G^M$ in Eq.~(\ref{G}) as a function of the detuning $\Delta_l$ and coupling strength $J$. The contour line corresponding to $G^M/\gamma_b= 1$ is marked by the black dashed line in Fig.~\ref{Fig7}. The region enclosed by the black dashed line indicates where the mechanical gain $G^M>\gamma_b$, which means the phonon number can be amplified. The peak of the mechanical gain with $G^M /\gamma_b= 2.5$ is obtained at $\Delta_l/\omega_b=0.68$ and $ J/\omega_b= 0.48$. This condition precisely corresponds to the resonance position of the supermode system. In this scenario, the lower supermode is sufficiently excited to the upper one, resulting in maximum population inversion and thus generating the strongest phonon laser.

In Fig.~\ref{Fig8}, we plot the isolation parameter $R_\Omega$ in Eq.~(\ref{R}) as a function of the detuning $\Delta_l$ and angular velocity $\Omega$ in forward input case. The black dashed, red dash-dotted, and blue solid lines respectively correspond to $R_\Omega=0, -5$, and $10$. The non-zero isolation parameter $R$ indicates the emergence of a nonreciprocal phenomenon in the system. In the forward-input case, the phonon laser is enhanced at $R_\Omega>0$, while the phonon amplification is suppressed at $R_\Omega<0$.
\begin{figure}[h]
\centerline{
\centering
\includegraphics[width=8cm, height=6cm, clip]{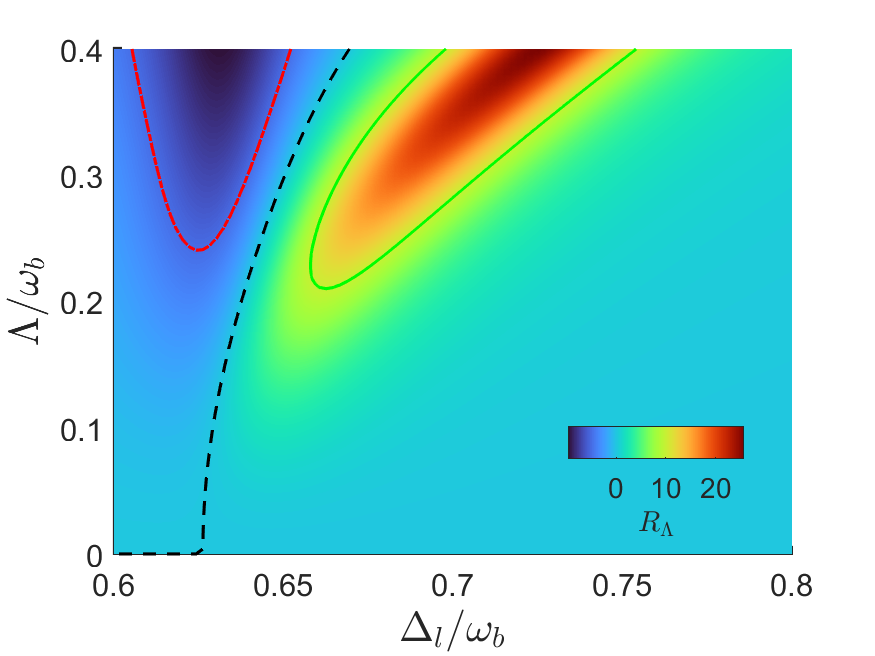}}
\caption{The isolation parameter $R_\Lambda$ in Eq.~(\ref{R}) is plotted as a function of the detuning $\Delta_l$ and squeezing strength $\Lambda$. The black dashed, red dash-dotted, and green solid lines respectively correspond to $R_\Lambda=0$, $R_\Lambda=-5$, and $R_\Lambda=10$. The parameter chosen is $\Omega=6\times 10^3$ Hz, while other parameters chosen are same as Fig.~\ref{Fig2}.}\label{Fig9}
\end{figure}

To study the influences of DOPA on the nonreciprocal phonon laser, the isolation parameter $R_\Lambda$ in Eq.~(\ref{R}) is plotted as a function of the detuning $\Delta_l$ and squeezing strength $\Lambda$ in Fig.~\ref{Fig9}. We plot the contour of different isolation parameters, where the green soild, black dashed, and red dash-dotted contour lines respectively correspond to $R_\Lambda=10, 0$, and $-5$. When the isolation parameter $R_\Lambda=0$, the influences of DOPA on the system is not considered. For $R_\Lambda > 0$, the directional quantum squeezing effect leads to a larger stimulated emitted phonon number. The reason for the phenomenon is that the squeezing effect enhances $J_s$, $g_s$, and $\varepsilon_s$. For the forward-input case, the strongest nonreciprocal phonon laser occurs at $\Delta_l = 0.72\omega_b$ when the squeezing strength is $\Lambda=0.4\omega_b$. Therefore, we can refer to Figs.~\ref{Fig8} and~\ref{Fig9} to identify suitable parameters for getting the enhanced nonreciprocal phonon laser.

\begin{figure}[h]
\centerline{
\centering
\includegraphics[width=9cm, height=7cm, clip]{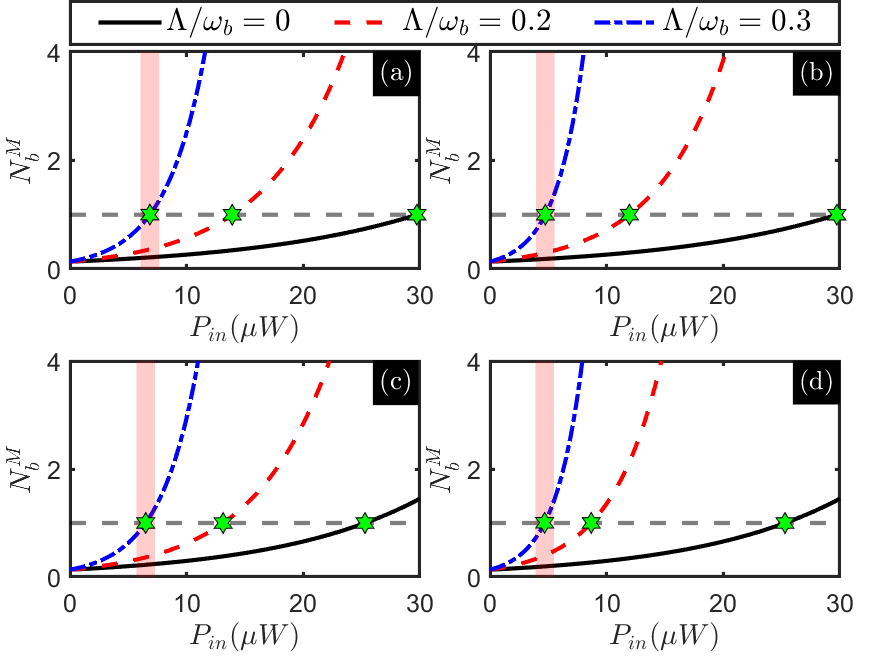}}
\caption{The stimulated emitted phonon number $N^M_b$ in Eq.~(\ref{Nb}) is plotted as a function of the pump power $P_{in}$. (a,c) DOPA is only driven by the pump field in $R_2$. (b,d) DOPA is only driven by the pump field in $R_1$. The parameters chosen are $\Omega=0$ with $\Delta_l=0.725\omega_b$ for (a,b) and $\Omega=6\times 10^3$ Hz with $\Delta_l=0.675\omega_b$ for (c,d), while other parameters chosen are same as Fig.~\ref{Fig2}.}\label{Fig15}
\end{figure}

To compare with previous works, we plot the relationship between the stimulated emitted phonon number and input power for DOPA in different resonators in Fig.~\ref{Fig15} (a) and (b). We find that when DOPA is only driven in $R_2$, the threshold power of the phonon laser is 6.85 $\mu$W, while for DOPA in resonator $R_1$, the threshold power decreases to 4.74 $\mu$W. This indicates that our scheme has a greater enhancement effect on the phonon laser. The reason is that when DOPA is in $R_1$, the squeezing effect not only leads to an exponential enhancement of the coupling strength $J_s$ but also enhances the optomechanical coupling $g_s$ and input amplitude $\varepsilon_s$, which is different from previous works \cite{Lu260312,Zhou2786}. Moreover, in Fig.~\ref{Fig15} (c) and (d), we find that Sagnac-Fizeau effects further reduce the threshold power of the phonon laser, resulting in a phonon laser with an ultra-low threshold power.
\section{nonreciprocal phonon laser in non-Markovian environments}\label{Sec4}
When the quantum system couples to external environments, the dynamics of the system influenced by the environments exhibits the dissipation or the backflow oscillation of the photon generated in the environments, which respectively correspond to the Markovian and non-Markovian regimes \cite{Vega015001,Breuer2002,Breuer021002,Caruso1203}. In this section, we investigate how non-Markovian effects affect the nonreciprocal phonon laser with DOPA fixed. For this purpose, we consider that two resonators interact with two non-Markovian environments consisting of a series of boson modes (eigenfrequencies $x_k$ and $n_k$). The total Hamiltonian including $\hat H_S$ in Eq.~(\ref{H0}) is changed to
\begin{equation}
\begin{aligned}
&{\hat H_N}=\hat H_S+\hat H_I, \hat H_I= \sum\limits_k {{x_k}\hat d_k^\dag } {\hat d_k} + \sum\limits_k ({W_k}{{\hat a}_1}\hat d_k^\dag\\&+ {\rm{H}}{\rm{.c}}.)
+ \sum\limits_k {{n_k}\hat e_k^\dag } {\hat e_k} + \sum\limits_k {({G_k}{{\hat a}_2}\hat e_k^\dag  + {\rm{H}}{\rm{.c}}.)}\label{Hsno},
\end{aligned}
\end{equation}where $\hat d_k$ ($\hat d_k^\dag$) and $\hat e_k$ ($\hat e_k^\dag$) denote the annihilation (creation) operators of two non-Markovian environments, respectively. $W_k$ and $G_k$ represent the coupling strengths. The Hamiltonian (\ref{Hsno}) via Bogoliubov transformation {(\ref{as}}) containing $\hat H_F$ in Eq.~(\ref{Heff}) is given by
\begin{eqnarray}
&{\hat H_N^{(1)}}&=\hat H_F+\hat H_I^{(1)}, \hat H_I^{(1)}= \sum\limits_k {{\tilde{x}_k}\hat d_k^\dag } {\hat d_k}+ \sum\limits_k {{\tilde{n}_k}\hat e_k^\dag } {\hat e_k}\nonumber\\
&+& \sum\limits_k ({W_k}{{\hat a}_s}\hat d_k^\dag + {\rm{H}}{\rm{.c}}.) + \sum\limits_k {({G_k}{{\hat a}_2}\hat e_k^\dag + {\rm{H}}{\rm{.c}}.)}\label{Hsno1},
\end{eqnarray}
where $\tilde{x}_k=x_k-\omega_e$ and $\tilde{n}_k=n_k-\omega_e$. With operator expectation values $a_s \equiv \langle \hat{a}_s \rangle$, $a_2 \equiv  \langle \hat{a}_2 \rangle$, $d_k \equiv \langle \hat{d}_k \rangle$, $e_k \equiv \langle \hat{e}_k \rangle$, and $b \equiv \langle \hat{b} \rangle$ in Eq.~(\ref{Hsno1}), we have (see Appendix \ref{DNME} for the detailed derivation)
\begin{eqnarray}
\dot{a}_s&=&i\Delta_s{a_s}+i g_s x_0({b+ {b}^*})a_s-i J_s a_2+\varepsilon_s e^{-i \Delta_1 t}-Y_1,\nonumber\\
\dot{a}_2&=&(i\Delta_l+i\Delta_{\rm{sag}}){a_2}-i J_s {a_s}-Y_2,\nonumber\\
\dot{b}&=&-(i\omega_b+\gamma_b)b+ i g_s x_0 {{a_s^*} }a_s,\label{eqano1}\\
\dot{Y_1}&=&- {\lambda _1}{Y_1} + \frac{1}{2}{\Gamma _1}{\lambda _1}{a_s},\,\,\dot{Y_2}=-{\lambda _2}{Y_2} + \frac{1}{2}{\Gamma _2}{\lambda _2}{a_2},\nonumber
\end{eqnarray}
where $Y_1=\int_{0}^{t}d{\tau}a_s(\tau)f_{1}(t-\tau)$ and $Y_2=\int_{0}^{t}d{\tau}a_2(\tau)f_{2}(t-\tau)$.
By defining ${a_ s } = {A_s}{e^{ - i{\Delta _1}t}}$ and ${a_ 2 } = {A_2}{e^{ - i{\Delta _1}t}}$, we arrive at
\begin{equation}
\begin{aligned}
 \ddot{A}_{s}=&  2i\Delta_1 \dot{A}_{s} + \Delta_1^2 A_{s}+
i\Delta_s ( \dot{A}_{s} - i\Delta_1 A_{s} )-i\Delta_1 \varepsilon_s \\
&+ i g_s x_s (\dot{A}_{s} - i\Delta_1 A_{s} ) + i g_s \dot{x}_s A_{s}  \\
&-\lambda_1 [ {\dot A_s} - i({\Delta _1}+\Delta_s+g_s x_s){A_s} + iJ_s A_2 ]\\
&+ \lambda_1\varepsilon_s - \frac{1}{2}\Gamma_1\lambda_1 A_{s}- iJ_s ( \dot{A}_2 - i\Delta_1 A_2 ),\\
 \ddot{A}_2=& 2i\Delta_1 \dot{A}_2+ \Delta_1^2 A_2 +
i(\Delta_l + \Delta_{\rm{sag}})(\dot{A}_2 - i\Delta_1 A_2 ) \\
&-\lambda_2 [ {\dot A_2} - i{\Delta _1}{A_2}- i(\Delta_l + \Delta_{\rm{sag}}) A_2 + iJ_s A_{s} ] \\
&- iJ_s ( \dot{A}_{s} - i\Delta_1 A_{s} ) - \frac{1}{2}\Gamma_2\lambda_2 A_2.\label{ddAs}
\end{aligned}
\end{equation}
According to Eq.~(\ref{ddAs}), we get the steady-state solutions
\begin{equation}
\begin{aligned}
A_s =& \frac{{-i(2i{h_2} + 4 \Delta _A)\varepsilon_s }}{{ 4 J_s^2 + [ h_1 -2i( {g_s}{x_s}+{\Delta _1} + {\Delta _s})]( h_2 -2i \Delta _A)}},\\
A_2 =& \frac{{-4i{J_s}\varepsilon_s }}{{ 4 J_s^2 + [ h_1 -2i( {g_s}{x_s}+{\Delta _1} + {\Delta _s})]( h_2 -2i \Delta _A)}},\label{Asno1}
\end{aligned}
\end{equation}
where ${h_v} = ({\Gamma _v}{\lambda _v})/({\lambda _v - i\Delta _1})$. With the definition of supermode operators $\hat a_+=(\hat a_s+\hat a_2)/\sqrt{2}$ and $\hat a_-=(\hat a_s-\hat a_2)/\sqrt{2}$, the Hamiltonian in Eq.~(\ref{Hsno1}) including $\hat H_s$ in Eq.~(\ref{He+}) can be written as
\begin{equation}
\begin{aligned}
{\hat H_N^{(2)}}=&\hat H_s+\hat H_I^{(2)},\\
\hat H_I^{(2)}=& \sum\limits_k {{\tilde{x}_k}\hat d_k^\dag } {\hat d_k} + \sum\limits_k {[{W_k}\frac{1}{\sqrt{2}}({{\hat a}_+ +{\hat a}_-})\hat d_k^\dag  + {\rm{H}}{\rm{.c}}.]}\\
&+ \sum\limits_k {{\tilde{n}_k}\hat e_k^\dag } {\hat e_k} + \sum\limits_k {[{G_k}\frac{1}{\sqrt{2}}({{\hat a}_+ -{\hat a}_-})\hat e_k^\dag  + {\rm{H}}{\rm{.c}}.]}.\label{Hsno+}
\end{aligned}
\end{equation}
With definitions $a_+ \equiv \langle \hat{a}_+ \rangle$, $a_- \equiv \langle \hat{a}_- \rangle$, and $b \equiv \langle \hat{b} \rangle$, Eq.~(\ref{Hsno+}) gives
\begin{equation}
\begin{aligned}
\dot{a}_+=&-{i \omega_+a_+ +\frac{i}{2}(g_s x_0 b-\Delta-\Delta_{\rm{sag}})a_-}\\
&+\frac{\varepsilon_s}{\sqrt{2}}e^{-i \Delta_1 t}-\frac{1}{2}Y_3-\frac{1}{2}Y_4,\\
\dot{a}_-=&-{i \omega_- a_- +\frac{i}{2}(g_s x_0 b^*-\Delta-\Delta_{\rm{sag}})a_+}\\
&+\frac{\varepsilon_s}{\sqrt{2}}e^{-i \Delta_1 t}-\frac{1}{2}Y_3+\frac{1}{2}Y_4,\\
\dot{Y_3}=&-\lambda_1 Y_3+\frac{1}{2}\Gamma_1 \lambda_1(a_+ +a_-),\\
\dot{Y_4}=&-\lambda_2 Y_4+\frac{1}{2}\Gamma_2 \lambda_2(a_+ -a_-),\\
\end{aligned}
\end{equation}
where $Y_3=\int_{0}^{t}d{\tau}[a_+(\tau)+a_-(\tau)]f_{1}(t-\tau)$ and $Y_4=\int_{0}^{t}d{\tau}[a_+(\tau)-a_-(\tau)]f_{2}(t-\tau)$. By introducing ${a_ + } = {A_+}{e^{ - i{\Delta _1}t}}$ and ${a_ - } = {A_-}{e^{ - i{\Delta _1}t}}$, we take $\lambda_2=\lambda_1$ to obtain
\begin{equation}
\begin{aligned}
\ddot{A}_+ = &2i\Delta_1 \dot{A}_+ + \Delta_1^2 A_+-i\omega_+ ( \dot{A}_+- i\Delta_1 A_+ )\\
& + \frac{i}{2} (g_s x_0 b - \Delta - \Delta_{\rm{sag}}) ( \dot{A}_- - i\Delta_1 A_- )\\
&+ \frac{i}{2} g_s x_0 \dot{b} {A}_--i\Delta_1 \frac{\varepsilon_s}{\sqrt{2}}- \frac{1}{4} [ A_+ (\Gamma_1 + \Gamma_2) ] \lambda_1 \\
&+ \lambda_1 \frac{\varepsilon_s}{\sqrt{2}}-\lambda_1(\dot{A}_+-i \Delta_1 A_+)\\
&+ \lambda_1 [ -i\omega_+ A_+ + \frac{i}{2} (g_s x_0 b - \Delta - \Delta_{\rm{sag}}) A_- ],\\
\ddot{A}_- = &2i\Delta_1 \dot{A}_- + \Delta_1^2 A_--i\omega_- ( \dot{A}_-- i\Delta_1 A_- )\\
& + \frac{i}{2} (g_s x_0 b^* - \Delta - \Delta_{\rm{sag}}) ( \dot{A}_+ - i\Delta_1 A_+ ) \\
&+ \frac{i}{2} g_s x_0 \dot{b}^* {A}_+-i\Delta_1 \frac{\varepsilon_s}{\sqrt{2}}- \frac{1}{4} [ A_- (\Gamma_1 + \Gamma_2) ] \lambda_1\\
&+ \lambda_1\frac{\varepsilon_s}{\sqrt{2}}-\lambda_1(\dot{A}_--i \Delta_1 A_-)\\
&+ \lambda_1 [ -i\omega_- A_- + \frac{i}{2} (g_s x_0 b^* - \Delta - \Delta_{\rm{sag}}) A_+ ],\label{A+}
\end{aligned}
\end{equation}
and steady-state solutions
\begin{equation}
\begin{aligned}
{A_ + } =& \frac{{i\varepsilon_s \left[ {-ih - \left( {\Delta  + 2{\Delta _1} + {\Delta _{\rm{sag}}} - b{x_0}{g_s} - 2{\omega _ - }} \right)} \right]}}{{\sqrt 2 \left[ {2\beta_2  - ih\left( {2{\Delta _1} + {\Delta _l} + {\Delta _s} + {\Delta _{\rm{sag}}}} \right)} \right]}},\\
{A_ - } =& \frac{{i\varepsilon_s \left[ {-ih - \left( {\Delta  + 2{\Delta _1} + {\Delta _{\rm{sag}}} - b^*{x_0}{g_s} - 2{\omega _ + }} \right)} \right]}}{{\sqrt 2 \left[ {2\beta_2  - ih\left( {2{\Delta _1} + {\Delta _l} + {\Delta _s} + {\Delta _{\rm{sag}}}} \right)} \right]}},
\end{aligned}
\end{equation}
where ${h} = ({\Gamma _0}{\lambda _1})/({\lambda _1 - i\Delta _1})$ and $\beta_2  = J_s^2 + h^2 / 4 - {\Delta _l}{\Delta _s} + (x_0^2g_s^2 n_b)/4 - \mathop{\rm Re}\nolimits(b){x_0}{g_s}\Delta/2 - \Delta _1^2 - {\Delta _1}{\Delta _l} - {\Delta _1}{\Delta _s} -\Delta_s \Delta_{\rm{sag}} -{\Delta _1}{\Delta _{\rm{sag}}} - \mathop{\rm Re}\nolimits(b){x_0}{\Delta _{\rm{sag}}}{g_s}/2$.
From Eq.~(\ref{Hsno+}), the dynamical equations of the system can be written as
\begin{equation}
\begin{aligned}
\dot{b}=&-(i \omega_b+\gamma_b)b+\frac{i}{2}g_s x_0 p,\\
\dot{p} = &-2iJ_s p + \frac{i}{2}(\Delta + \Delta_{\rm{sag}} - g_s x_0 b)\delta{n} \\&+ \frac{1}{\sqrt{2}}(\varepsilon_s e^{i \Delta_1 t}a_+ + \varepsilon_s e^{-i \Delta_1 t}a_{-}^*) \\
&- \frac{1}{\sqrt{2}}(a_{-}^* Y_3 + a_+Y_3^*)- \frac{1}{\sqrt{2}}(a_{-}^* Y_4 - a_+Y_4^*),\\
\dot{Y}_3=& -\lambda_1 Y_3 + \frac{1}{2}\Gamma_1 \lambda_1 (a_+ + a_-),\\
\dot{Y}_4=& -\lambda_1 Y_4 + \frac{1}{2}\Gamma_2 \lambda_1 (a_+ - a_-).\label{b2}
\end{aligned}
\end{equation}
According to Eq.~(\ref{b2}), we get
\begin{equation}
\begin{aligned}
 \dot{p}=& - p \tilde{h}\Gamma_0-2iJ_s p + \frac{i}{2}(\Delta + \Delta_{\mathrm{\rm{sag}}} - g_s x_0 b)\delta{n} \\
 &+ \frac{1}{\sqrt{2}}(\varepsilon_s A_+ + \varepsilon_s A_{-}^*),\label{p2}
\end{aligned}
\end{equation}
which leads to the steady-state solution
\begin{equation}
\begin{aligned}
p = \frac{-i\sqrt{2}(A_{-}^* + A_{+})\varepsilon_s + \delta{n} (\Delta + \Delta_{\rm{sag}} - b x_0 g_s)}{4J_s - 2i\tilde{h}\Gamma_0  - 2\omega_b},\label{p4}
\end{aligned}
\end{equation}
where $\tilde{h}=\lambda_1^2/(\Delta_1^2+\lambda_1^2)$. From Eqs.~(\ref{b2}) and~(\ref{p4}), the dynamical equation of $b$ becomes
\begin{equation}
\begin{aligned}
\dot b = ( - i{\omega _b} - i{\omega ^{Non}} + G^{Non} - {\gamma _b})b + F^{Non},\label{b4}
\end{aligned}
\end{equation}
with the non-Markovian mechanical gain
\begin{eqnarray}
G^{Non}&=&G^{Non}_0+G^{Non}_1+G^{Non}_2,\nonumber\\
{G^{Non}_0} &= &\frac{{x_0^2{\tilde{h}\Gamma _0}\delta n g_s^2}}{{4[ {\tilde{h}^2\Gamma _0^2 + {{( {2{J_s} - {\omega _b}} )}^2}} ]}},\label{Gno}\\
{G^{Non}_1}& = &  \mathop{\rm Re}\nolimits\{\frac{{ x_0^2[-2 \tilde{h} \beta_2 \Gamma_0 -h \Delta_g( {2{J_s}-{\omega _b}} )]{\varepsilon_s ^2}g_s^2}}{{4( {{4\beta_2 ^2} + {h^2}{\Delta_g^2}} )[ {\tilde{h}^2\Gamma _0^2 + {{( { - 2{J_s} + {\omega _b}} )}^2}} ]}}\},\nonumber\\
{G^{Non}_2} &= &  \mathop{\rm Re}\nolimits\{\frac{{ x_0^2[2 \tilde{h} \beta_2^* \Gamma_0 -h^* \Delta_g( {2{J_s}-{\omega _b}} )]{\varepsilon_s ^2}g_s^2}}{{4( {{{4\beta_2^*} ^2} + {{h^*}^2}{\Delta_g^2}} )[ {\tilde{h}^2\Gamma _0^2 + {{( { - 2{J_s} + {\omega _b}} )}^2}} ]}}\},\nonumber
\end{eqnarray}
where the first term of $G^{Non}$ is proportional to the population inversion $\delta n$ with non-Markovian effects. In non-Markovian regimes, $G^{Non}_0$, $G^{Non}_1$, and $G^{Non}_2$ depend on $\lambda_1$, which is different that in Eq.~(\ref{G}). For different $\lambda_1$, different mechanical gains $G^{Non}$ can be obtained. This means that we can tune the nonreciprocal phonon laser by the environmental spectrum width $\lambda_1$ with DOPA fixed. We do not give the expressions of $\omega ^{Non}$ and $F^{Non}$ in Eq.~(\ref{b4}) because they do not effect the phonon laser threshold power. In non-Markovian environments, the threshold condition under which the phonon mode can be amplified satisfies $G^{Non} = \gamma_b$. In this case, the stimulated emitted phonon number is
\begin{equation}
\begin{aligned}
{N_b^{Non}} = \exp [2(G^{Non} - \gamma_b)/\gamma_b],\label{Nbnon}
\end{aligned}
\end{equation}
which characterizes the performance of the phonon laser with non-Markovian effects.
\section{Influences of non-Markovian effects on nonreciprocal phonon laser}\label{Sec5}
In numerical simulations, we select experimentally feasible parameters with $n=1.48$, $\omega_1/{2 \pi}=\omega_2/{2 \pi}=30.7$ THz, $r_1=33$ $\mathrm{\mu{m}}$, $Q_1=9.7\times10^7$, $r_2=4.75$ mm, $Q_2=3\times10^7$, $m=50$ ng, $\gamma_b=0.24$ MHz, $J=0.48\omega_b$, $\omega_b / 2\pi=23.4$ MHz, $\lambda_1=0.5\omega_b$, $\Delta_1=-0.2\omega_b$, $P_{in}=10$ $\mu$W, and $\Lambda=0.3\omega_b$ \cite{Jiang064037,Grudinin083901}.
\begin{figure}[t]
\centerline{
\centering
\includegraphics[width=8cm, height=6cm, clip]{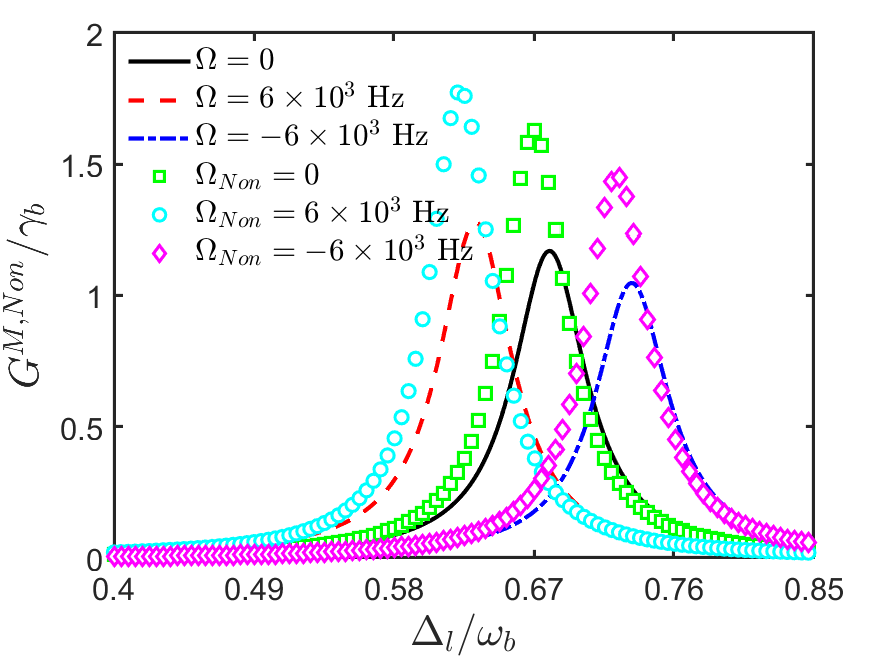}}
\caption{The mechanical gains $G^M$ and $G^{Non}$ determined by Eqs.~({\ref{G}}) and~({\ref{Gno}}) are plotted as a function of the detuning $\Delta_l$. The square, circle, and diamond denote that non-Markovian effects are considered for the phonon laser with $\lambda_1=0.5\omega_b$. The parameters chosen are $n=1.48$, $\omega_1/{2 \pi}=\omega_2/{2 \pi}=30.7$ THz, $r_1=33$ $\mathrm{\mu{m}}$, $Q_1=9.7\times10^7$, $r_2=4.75$ mm, $Q_2=3\times10^7$, $m=50$ ng, $\gamma_b=0.24$ MHz, $J=0.48\omega_b$, $\omega_b / 2\pi=23.4$ MHz, $\Delta_1=-0.2\omega_b$, $\Lambda=0$, $P_{in}=10$ $\mu$W, and $\lambda_1=0.5\omega_b$.}\label{Fig10}
\end{figure}
In Fig.~\ref{Fig10}, the mechanical gains $G^M$ and $G^{Non}$ in Eqs.~({\ref{G}}) and~({\ref{Gno}}) as a function of the detuning $\Delta_l$ are plotted. The black solid, red dashed, and blue dash-dotted lines respectively correspond to $\Omega=$ 0, $6\times10^3$ Hz, and $-6\times10^3$ Hz under the Markovian approximation. Note that DOPA is not considered in Fig.~\ref{Fig10}. The spinning of the resonator leads to different frequency shifts for the mechanical gain $G^{Non}$, which is similar to the Markovian case. Compared to the Markovian approximation, the value of $G^{Non}$ with non-Markovian effects is larger than the mechanical gain $G^M$, indicating that non-Markovian effects have enhancing influences on the phonon laser. The enhancement originates from the excitation backflow obtained through the interaction between resonators and non-Markovian environments. We can realize a phonon laser with enhanced performance with non-Markovian effects beyond the Markovian approximation.
\begin{figure}[h]
\centerline{
\centering
\includegraphics[width=8cm, height=6cm, clip]{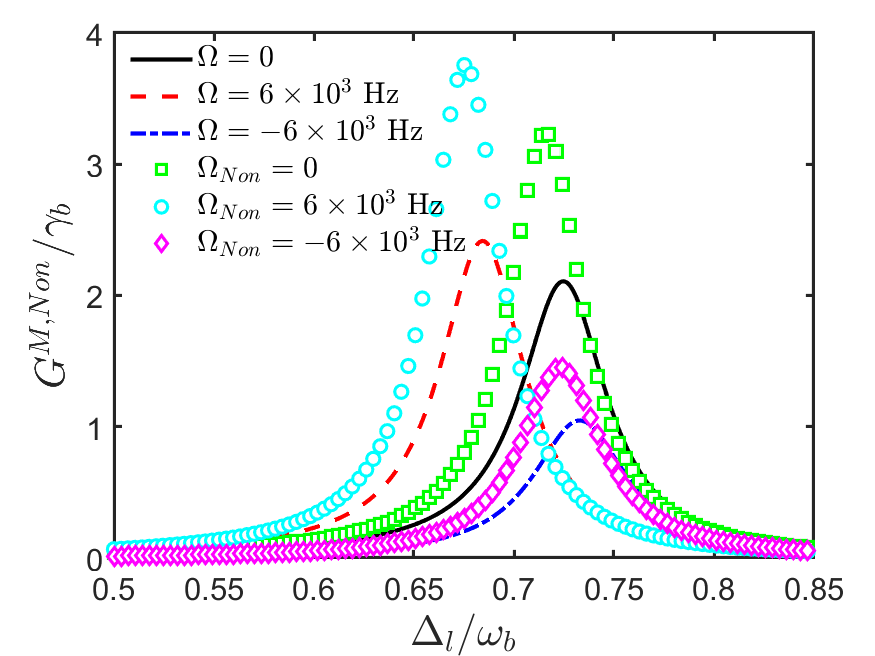}}
\caption{The mechanical gains $G^M$ and $G^{Non}$ determined by Eqs.~({\ref{G}}) and~({\ref{Gno}}) are plotted as a function of the detuning $\Delta_l$ with squeezing strength $\Lambda=0.3\omega_b$. The square, circle, and diamond denote that non-Markovian effects are considered for the phonon laser with $\lambda_1=0.5\omega_b$. The black solid, red dashed, and blue dash-dotted lines correspond to the phonon laser in the Markovian environment. The parameters chosen are same as Fig.~\ref{Fig10}.}\label{Fig11}
\end{figure}

In Fig.~\ref{Fig11}, we plot the mechanical gains $G^M$ and $G^{Non}$ in Eqs.~({\ref{G}}) and~({\ref{Gno}}) as a function of the detuning $\Delta_l$ with DOPA fixed. In contrast to Fig.~\ref{Fig10}, we consider the influences of DOPA on the mechanical gain in Fig.~\ref{Fig11}. It is observed that DOPA leads to an enhancement in the mechanical gain. For $\Omega=6 \times 10^3$ Hz, the mechanical gain $G^{Non}$ increases to 3.76$\gamma_b$ and mechanical gain $G^M$ arrives at 2.42$\gamma_b$, which are larger than those cases without DOPA. Moreover, we find that non-Markovian effects still significantly enhance the mechanical gain, which is independent of whether DOPA is considered.
\begin{figure}[h]
\centerline{
\centering
\includegraphics[width=8cm, height=6cm, clip]{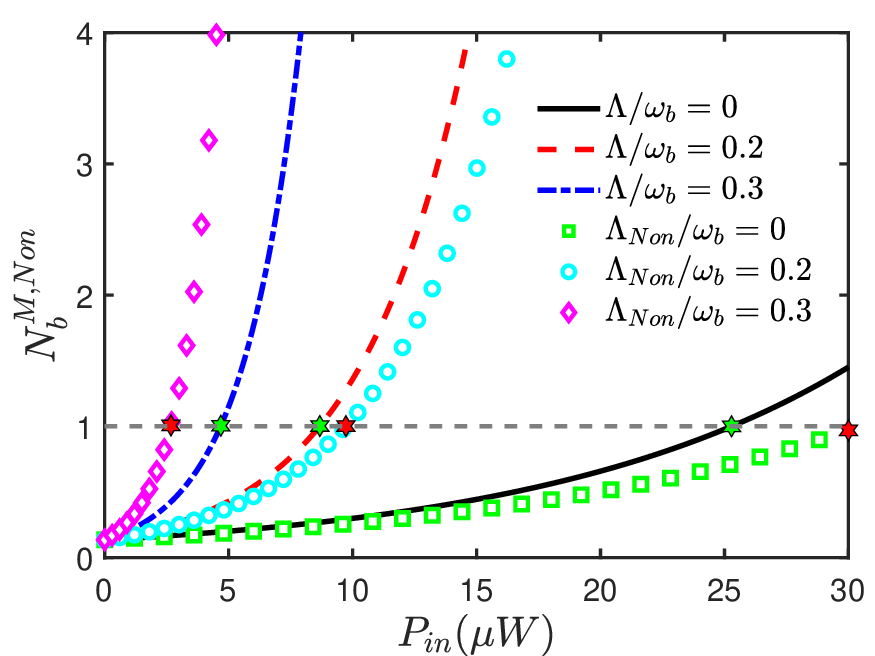}}
\caption{The stimulated emitted phonon numbers $N^M_b$ and $N_b^{Non}$ determined by Eqs.~({\ref{Nb}}) and~({\ref{Nbnon}}) are plotted as a function of the pump power $P_{in}$. The black solid, red dashed, and blue dash-dotted lines respectively correspond to squeezing strengths $\Lambda/\omega_b=0$, $\Lambda/\omega_b=0.2$, and $\Lambda/\omega_b=0.3$ in the Markovian environment. The square, circle, and diamond denote that non-Markovian effects are considered for the phonon laser with $\lambda_1=0.5\omega_b$. The six-pointed stars mark the threshold power, which is determined by using the threshold condition $G^{M,Non}=\gamma_b$ in Eqs.~({\ref{Nb}}) and~({\ref{Nbnon}}). The parameters chosen are $\Omega=6\times10^3$ Hz and $\Delta_l=0.675\omega_b$, while other parameters chosen are same as Fig.~\ref{Fig10}.}\label{Fig12}
\end{figure}

Figure~\ref{Fig12} shows the relationship between the stimulated emitted phonon numbers and pump power $P_{in}$. We find that non-Markovian effects reduce the threshold power of the phonon laser under the same conditions. Compared to the Markovian case, the threshold pump power can be decreased from 4.71 $\mu$W to 2.7 $\mu$W with $\Lambda/\omega_b=0.3$, which implies that the phonon laser can be realized at a lower power level. However, for squeezing strength $\Lambda/\omega_b=0.2$, non-Markovian effects suppress phonon number under stimulated emission, where the threshold power increases from 8.68 $\mu$W to 9.76 $\mu$W. Therefore, we can realize the nonreciprocal phonon laser with the lower threshold power and  higher isolation ratio by introducing non-Markovian effects.
\begin{figure}[t]
\centerline{
\centering
\includegraphics[width=8cm, height=6cm, clip]{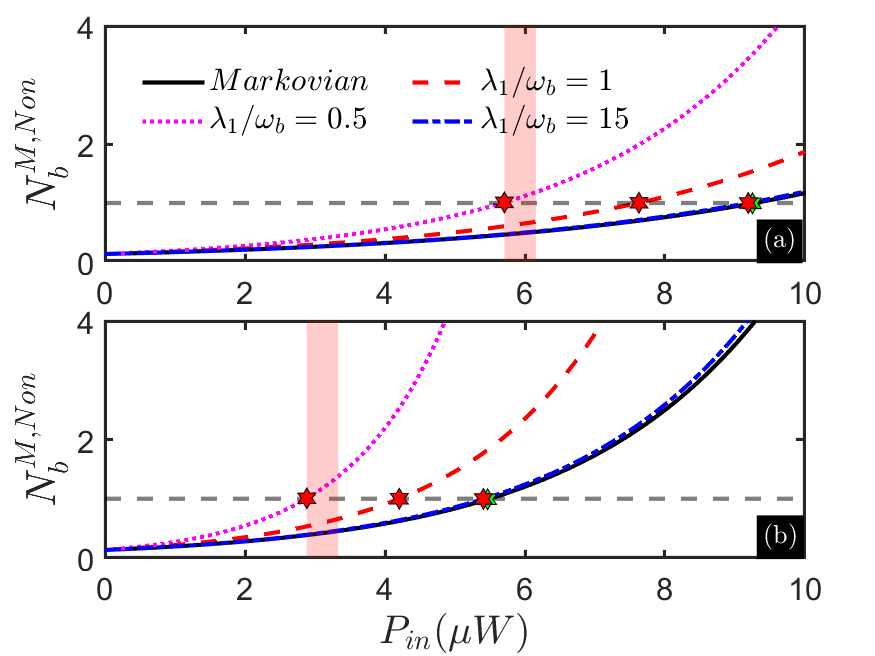}}
\caption{The stimulated emitted phonon numbers $N^M_b$ and $N_b^{Non}$ determined by Eqs.~({\ref{Nb}}) and~({\ref{Nbnon}}) are plotted as a function of the pump power $P_{in}$ with Markovian (black soild line) and non-Markovian regimes in (a) $\Lambda/\omega_b=0$ and (b) $\Lambda/\omega_b=0.45$. The pink dotted, red dashed, and blue dash-dotted lines respectively correspond to non-Markovian environmental spectrum widths $\lambda_1/\omega_b=0.5$, $\lambda_1/\omega_b=1$, and $\lambda_1/\omega_b=15$. The parameters chosen are $\Delta_l=0.62\omega_b$ for (a) and $\Delta_l=0.73\omega_b$ for (b), while other parameters chosen are same as Fig.~\ref{Fig12}.}\label{Fig13}
\end{figure}

In order to verify the validity of our method, we simulate our system under the large environmental spectrum width scenario. In Fig.~\ref{Fig13}(a), the nonreciprocal phonon laser exhibits significant differences between Markovian and non-Markovian cases with the environmental spectrum width $\lambda_1 = 0.5\omega_b$. As the spectrum width of the environment increases, the threshold pump power of the phonon laser is continuously rising, which is attributed to the enhancement of the mechanical gain by non-Markovian effects. This means the smaller environmental spectral width leads to lower threshold power of the phonon laser. Moreover, under the same input power, more phonon numbers are stimulated to emit with the smaller environmental spectrum width, as indicated by the shaded region in Fig.~{\ref{Fig13}}(a). For the sake of clarity, we separately study two cases with the Markovian approximation and non-Markovian limit under the environmental spectrum width $\lambda_1 = 15\omega_b$. In this case, we find that the stimulated emitted phonon number is consistent with the Markovian approximation, which means that the system returns to the Markovian regimes. In Fig.~{\ref{Fig13}}(b), we plot the stimulated emitted phonon numbers as a function of the pump power for the system with DOPA fixed. By comparing Fig.~{\ref{Fig13}}(a) and (b), we find that DOPA leads to a reduction in the threshold power for the phonon laser in non-Markovian environments. The pump threshold power required to realize phonon amplification is 2.88 $\mu$W when DOPA is considered, whereas it increases to 5.7 $\mu$W without DOPA. This indicates that the enhancement of the phonon laser arises from the dual influences by DOPA and non-Markovian effects in our scheme.
\section{Conclusion}\label{Sec6}
In summary, we have demonstrated the realization of the enhanced nonreciprocal phonon laser in a coupled cavity system composed of nonlinear optomechanical resonator and DOPA under the Markovian approximation. By spinning the resonator, Sagnac-Fizeau shifts lead to alterations in the mechanical gain and threshold power, inducing nonreciprocal phonon laser. Moreover, we extended the nonreciprocal phonon laser to non-Markovian regimes, where the results obtained under the Markovian approximation are consistent with those in the non-Markovian limit. We found that non-Markovian memory effects can enhance the mechanical gain and contribute to a reduction of the phonon laser threshold power with DOPA fixed. These results show low-power non-Markovian mechanical isolation and offer a platform for lower power cavity optomechanical control.

The enhanced nonreciprocal phonon laser can realize directional emission and amplification of phonons, which is very important for acoustic sensing or computing. In this case, our discussion about non-Markovian effects provides a reference for the enhancement of the nonreciprocal phonon laser.  As an outlook, we will explore general phonon laser models with non-rotating-wave interactions between resonators and atom under ultrastrong coupling~\cite{Chen033603,Chen043711,Yan180401}, as well as between resonators and environments $\sum_kV(\hat{z}+\hat{z}^{\dagger})(\hat{c}_k+{\hat{c}}^{\dagger}_k)$ \cite{Shen033805,Shen042121}, anisotropic non-rotating-wave ones $\sum_{k}[\chi_k(\hat{c}^{\dagger}_k\hat{z}+\hat{c}_k\hat{z}^{\dagger})+\zeta_k(\hat{c}_k\hat{z}
+\hat{c}^{\dagger}_k\hat{z}^{\dagger})]$ \cite{Xie021046,Chen043708, Nakajima363,Ai042116,Zheng559}, and even beyond the anisotropic non-rotating-wave situations, i.e, all couplings including systems and environments might be of the form $\sum_{m,k}(L_{m,k}\hat{I}_m\hat{O}_k^\dagger+ L_{m,k}^*\hat{I}_m^\dagger\hat{O}_k+ G_{m,k}\hat{I}_m\hat{O}_k+ G_{m,k}^*\hat{I}_m^\dagger\hat{O}_k^\dagger)$, where $\hat{I}_m^\dagger(\hat{I}_m)$ and $\hat{O}_k^\dagger(\hat{O}_k)$ correspond to the creation (annihilation) operators for systems, while $L_{m,k}$ and $G_{m,k}$ respectively denote the strengths of rotating wave and non-rotating-wave couplings.
\section*{ACKNOWLEDGMENTS}

This work was supported by Science and Technology Development Plan Project of Jilin Province (Grant No.~20250102007JC), National Natural Science Foundation of China (NSFC) under Grants No.~12274064 and No. 12374333, Shandong Provincial Natural Science Foundation under Grant No. ZR2021MA036, and the Hakubi Projects of RIKEN.
\section*{DATA AVAILABILITY}
The data that support the findings of this article are not publicly available. The data are available from the authors upon reasonable request.

\appendix
\section{Derivation of the Supermode Hamiltonian}
\label{DSH}
Here, we provide the detailed transformation from the effective Hamiltonian in Eq.~(\ref{Heff}) to the supermode representation given in Eq.~(\ref{He+}). Starting from the effective Hamiltonian in Eq.~(\ref{Heff}) for the forward-input case, we introduce the supermode operators \(\hat{a}_+ = (\hat{a}_s + \hat{a}_2)/\sqrt{2}\) and \(\hat{a}_- = (\hat{a}_s - \hat{a}_2)/\sqrt{2}\), which satisfy \([\hat{a}_\pm,\hat{a}_\pm^\dagger]=1\) and \([\hat{a}_+,\hat{a}_-^\dagger]=0\). The inverse transformations are \(\hat{a}_s = (\hat{a}_+ + \hat{a}_-)/\sqrt{2}\) and \(\hat{a}_2 = (\hat{a}_+ - \hat{a}_-)/\sqrt{2}\). Substituting these into Eq.~(\ref{Heff}) gives
\begin{equation}
\begin{aligned}
-\Delta_s \hat{a}_s^\dagger \hat{a}_s =& -\frac{\Delta_s}{2}(\hat{a}_+^\dagger + \hat{a}_-^\dagger)(\hat{a}_+ + \hat{a}_-)\\
=& -\frac{\Delta_s}{2}(\hat{a}_+^\dagger \hat{a}_+ + \hat{a}_+^\dagger \hat{a}_-\\& + \hat{a}_-^\dagger \hat{a}_+ + \hat{a}_-^\dagger \hat{a}_-),\\
-(\Delta_l + \Delta_{\text{sag}})\hat{a}_2^\dagger \hat{a}_2 =& -\frac{\Delta_l + \Delta_{\text{sag}}}{2}(\hat{a}_+^\dagger - \hat{a}_-^\dagger)(\hat{a}_+ - \hat{a}_-)\\
=& -\frac{\Delta_l + \Delta_{\text{sag}}}{2}(\hat{a}_+^\dagger \hat{a}_+ - \hat{a}_+^\dagger \hat{a}_-\\& - \hat{a}_-^\dagger \hat{a}_+ + \hat{a}_-^\dagger \hat{a}_-),\\
\hat{a}_s^\dagger \hat{a}_2 =& \frac{1}{2}(\hat{a}_+^\dagger + \hat{a}_-^\dagger)(\hat{a}_+ - \hat{a}_-)\\
=& \frac{1}{2}(\hat{a}_+^\dagger \hat{a}_+ - \hat{a}_+^\dagger \hat{a}_- + \hat{a}_-^\dagger \hat{a}_+ - \hat{a}_-^\dagger \hat{a}_-),\\
\hat{a}_2^\dagger \hat{a}_s =& \frac{1}{2}(\hat{a}_+^\dagger - \hat{a}_-^\dagger)(\hat{a}_+ + \hat{a}_-)\\
=& \frac{1}{2}(\hat{a}_+^\dagger \hat{a}_+ + \hat{a}_+^\dagger \hat{a}_- - \hat{a}_-^\dagger \hat{a}_+ - \hat{a}_-^\dagger \hat{a}_-).
\end{aligned}
\end{equation}
Using the relation \(({\Delta_l + \Delta_{\text{sag}} - \Delta_s})/{2} = ({\Delta + \Delta_{\text{sag}}})/{2}\) with \(\Delta = \Delta_l - \Delta_s\), we obtain
\begin{equation}
\begin{aligned}
\mathcal{H}=&\omega_+ \hat{a}_+^\dagger \hat{a}_+ + \omega_- \hat{a}_-^\dagger \hat{a}_- + \frac{\Delta + \Delta_{\text{sag}}}{2} (\hat{a}_+^\dagger \hat{a}_- + \hat{a}_-^\dagger \hat{a}_+)\\&+J_s (\hat{a}_+^\dagger \hat{a}_+ - \hat{a}_-^\dagger \hat{a}_-).\label{HH}
\end{aligned}
\end{equation}
For the optomechanical coupling term \(-g_s x_0 \hat{a}_s^\dagger \hat{a}_s (\hat{b}^\dagger + \hat{b})\), we have
\begin{equation}
\begin{aligned}
-g_s x_0 \hat{a}_s^\dagger \hat{a}_s (\hat{b}^\dagger + \hat{b})=&-\frac{g_s x_0}{2}(\hat{a}_+^\dagger \hat{a}_+ + \hat{a}_-^\dagger \hat{a}_- \\&+ \hat{a}_+^\dagger \hat{a}_- + \hat{a}_-^\dagger \hat{a}_+)(\hat{b}^\dagger + \hat{b}).\label{opt}
\end{aligned}
\end{equation}
In the rotating frame with respect to $\hat H_0 = \omega_+ \hat{a}_+^\dagger \hat{a}_+ + \omega_- \hat{a}_-^\dagger \hat{a}_- + \omega_b \hat{b}^\dagger \hat{b}$, Eq.~\eqref{opt} becomes
\begin{align*}
-\frac{g_s x_0}{2} [ & a_+^\dagger a_- b \, e^{i(2J_s-\omega_b)t} + a_+ a_-^\dagger b^\dagger \, e^{-i(2J_s-\omega_b)t} \\
& + a_+^\dagger a_- b^\dagger \, e^{i(2J_s+\omega_b)t} + a_+ a_-^\dagger b \, e^{-i(2J_s+\omega_b)t} \\
& + (a_+^\dagger a_+ + a_-^\dagger a_-)(b^\dagger e^{i\omega_b t} + b e^{-i\omega_b t}) ].
\end{align*}
Under the rotating-wave approximation, which requires $2J_s+\omega_b,\omega_b \gg |2J_s-\omega_b|$ \cite{Jiang064037, Lu260312}, the terms $a_+^\dagger a_- b^\dagger \, e^{i(2J_s+\omega_b)t}$, $a_+ a_-^\dagger b \, e^{-i(2J_s+\omega_b)t}$, and $(a_+^\dagger a_+ + a_-^\dagger a_-)(b^\dagger e^{i\omega_b t} + b e^{-i\omega_b t})$ are therefore rapidly oscillating compared with the near-resonant terms $\hat{a}_+^\dagger \hat{a}_- \hat{b}$ and $\hat{b}^\dagger \hat{a}_-^\dagger \hat{a}_+$, which oscillate at the much lower frequency $|2J_s - \omega_b|$. The fast-oscillating terms are neglected, while the near-resonant terms are retained. The optomechanical coupling term in Eq.~\eqref{opt} simplifies to
\begin{equation}
\begin{aligned}
-\frac{g_s x_0}{2} (\hat{a}_+^\dagger \hat{a}_- \hat{b} + \hat{b}^\dagger \hat{a}_-^\dagger \hat{a}_+).
\end{aligned}
\end{equation}
The driving term \(i\varepsilon_s (\hat{a}_s^\dagger e^{-i\Delta_1 t} - \hat{a}_s e^{i\Delta_1 t})\) in Eq.~\eqref{Heff}  becomes
\begin{equation}
\begin{aligned}
i\varepsilon_s (\hat{a}_s^\dagger e^{-i\Delta_1 t} - \hat{a}_s e^{i\Delta_1 t})
=& \frac{i\varepsilon_s}{\sqrt{2}} [ (\hat{a}_+^\dagger + \hat{a}_-^\dagger) e^{-i\Delta_1 t} \\&- (\hat{a}_+ + \hat{a}_-) e^{i\Delta_1 t} ].
\end{aligned}
\end{equation}
Finally, we have
\begin{equation}
\begin{aligned}
\hat{H}_s =& \omega_+ \hat{a}_+^\dagger \hat{a}_+ + \omega_- \hat{a}_-^\dagger \hat{a}_- + \frac{\Delta + \Delta_{\text{sag}}}{2} (\hat{a}_+^\dagger \hat{a}_- + \hat{a}_-^\dagger \hat{a}_+) \\
&+ \omega_b \hat{b}^\dagger \hat{b} - \frac{g_s x_0}{2} (\hat{a}_+^\dagger \hat{a}_- \hat{b} + \hat{b}^\dagger \hat{a}_-^\dagger \hat{a}_+) \\
&+ \frac{i\varepsilon_s}{\sqrt{2}} [ (\hat{a}_+^\dagger + \hat{a}_-^\dagger) e^{-i\Delta_1 t} - (\hat{a}_+ + \hat{a}_-) e^{i\Delta_1 t} ],
\end{aligned}
\end{equation}
which is Eq.~(\ref{He+}).
\section{Derivation of non-Markovian equation Eq.~(\ref{eqano1})}\label{DNME}
With operator expectation values $a_s \equiv \langle \hat{a}_s \rangle$, $a_2 \equiv  \langle \hat{a}_2 \rangle$, $d_k \equiv \langle \hat{d}_k \rangle$, $e_k \equiv \langle \hat{e}_k \rangle$, and $b \equiv \langle \hat{b} \rangle$ in Eq.~(\ref{Hsno1}), Heisenberg equations give
\begin{equation}
\begin{aligned}
\dot{a}_s=&i\Delta_s{a_s}+i g_s x_0({b+ {b}^*})a_s-i J_s a_2+\varepsilon_s e^{-i \Delta_1 t}\\
&-i\sum_{k}W_{k}^{*}d_{k},\\
\dot{a}_2=&(i\Delta_l+i\Delta_{\rm{sag}}){a_2}-i J_s {a_s}-i\sum_{k}G_{k}^{*}e_{k},\\
\dot{b}=&-(i\omega_b+\gamma_b)b- i g_s x_0 {{ a_s^*} }a_s,\\
\dot{d}_k=&-i\tilde{x}_{k}d_{k}-iW_{k}a_s,\,\,\dot{e}_k=-i\tilde{n}_{k}e_{k}-iG_{k}a_2.\label{eqano}
\end{aligned}
\end{equation}Solving Eq.~\eqref{eqano}, solutions of the environments parts for $t\geq0$ are obtained by
\begin{equation}
\begin{aligned}
d_{k}&=d_{k}(0)e^{-i\tilde{x}_{k}t}-iW_{k}\int_{0}^{t}d{\tau}a_s(\tau)e^{-i\tilde{x}_{k}(t-\tau)},\\
e_{k}&=e_{k}(0)e^{-i\tilde{n}_{k}t}-iG_{k}\int_{0}^{t}d{\tau}a_2(\tau)e^{-i\tilde{n}_{k}(t-\tau)},\label{dotdef}
\end{aligned}
\end{equation}
where the first terms on the right-hand side of Eq.~\eqref{dotdef} denote the free-propagating parts of environment fields, while the influences of environments on the resonator are described by the second terms. By substituting Eq.~\eqref{dotdef} into Eq.~\eqref{eqano}, we get non-Markovian integro-differential equations
\begin{equation}
\begin{aligned}
\dot{a}_s  =& i\Delta_s{a_s}+i g_s x_0({b+ {b}^*})a_s-i J_s a_2+\varepsilon_s e^{-i \Delta_1 t}\\&-K_{1} -\int_{0}^{t}d{\tau}a_s(\tau)f_{1}(t-\tau), \\
\dot{a}_2  =& (i\Delta_l+i\Delta_{\rm{sag}}){a_2}-i J_s {a_s}-K_{2}\\
&-\int_{0}^{t}d{\tau}a_2(\tau)f_{2}(t-\tau),\label{docta}
\end{aligned}
\end{equation}where $K_{1}=i\sum_{k}W_{k}^{*}d_{k}(0)e^{-i\tilde{x}_{k}t}=\int_{-\infty}^{+\infty}d{\tau}\kappa_{b}^{*}(t-\tau)a_s^{in}(\tau)$ and $K_{2}=i\sum_{k}G_{k}^{*}e_{k}(0)e^{-i\tilde{n}_{k}t}=\int_{-\infty}^{+\infty}d{\tau}\kappa_{2}^{*}(t-\tau)a_2^{in}(\tau)$. The resonators interact with the incoming and outgoing modes of environments at both ends, where the expectation values of the input field operators are defined as $a_s^{in}(t)=-\sum_{k}e^{-i\tilde{x}_{k}t}d_{k}(0)/{\sqrt{2\pi}}$ and $a_2^{in}(t)=-\sum_{k}e^{-i\tilde{n}_{k}t}e_{k}(0)/{\sqrt{2\pi}}$. Moreover, the impulse response functions are expressed as $\kappa_{1}(t)=i\sum_{k}e^{i\tilde{x}_{k}t}W_{k}/{\sqrt{2\pi}}$ and $\kappa_{2}(t)=i\sum_{k}e^{i\tilde{n}_{k}t}G_{k}/{\sqrt{2\pi}}$, or in the continuum
\begin{equation}
\begin{aligned}
\kappa_{1}(t-{\tau})=&\frac{i}{\sqrt{2\pi}}\int_{-\infty}^{+\infty}{e^{i(\omega-\omega_e){(t-\tau)}}}W(\omega)d\omega,\\
\kappa_{2}(t-{\tau})=&\frac{i}{\sqrt{2\pi}}\int_{-\infty}^{+\infty}{e^{i(\omega-\omega_e){(t-\tau)}}}G({\omega})d\omega,\label{kappac}
\end{aligned}
\end{equation}where we have made replacements $W_k\rightarrow{W(\omega)}$ and $G_k\rightarrow{G(\omega)}$. The correlation functions $f_1(t)$ and $f_2(t)$ in Eq.~(\ref{docta}) describe the non-Markovian fluctuation-dissipation relationship of the environments and are given by $f_{v}(t)=\int_{-\infty}^{+\infty}{J_{v}(\omega)e^{-i(\omega-\omega_e){t}}}d\omega
=\int_{-\infty}^{+\infty}\kappa_v(\tau-\zeta)\kappa^*_v(t-\zeta)d\zeta\,(v=1,2)$, where $J_{1}(\omega)=|W(\omega)|^{2}$ and $J_{2}(\omega)=|G(\omega)|^{2}$ represent spectral densities of two environments. Here, impulse response function can be set as
$\kappa_{v}(t)=i\lambda_{v}\sqrt{\Gamma_{{v}}}e^{\lambda_{v} t}\theta(-t)$, where $\theta(t)$ is the unit step function ($\theta(t)$ = 1 for $t\geq{0}$ otherwise $\theta(t)=0$). It leads to correlation function $f_{v}(t)=\lambda_{v}\Gamma_{{v}}e^{-\lambda_{v}|t|}/2$. Applying the Fourier transform to Eq.~\eqref{kappac}, we obtain $W(\omega)=\sqrt{{\Gamma_{1}}/{2\pi}}\lambda_{1}/[{\lambda_{1}-i(\omega-\omega_e)}]$ and $G(\omega)=\sqrt{{\Gamma_{2}}/{2\pi}}{\lambda_{2}}/[{\lambda_{2}-i(\omega-\omega_e)}]$, where $\lambda_{v}$ is the environmental spectrum width, while $\Gamma_{v}$ is defined as the disspation of the environmental modes acting on the cavities. Thus the spectral density of the environments reads $J_{v}(\omega)={\Gamma_{v}}{\lambda_{v}^{2}}/\{{2\pi}[{\lambda_{v}^{2}+(\omega-\omega_e)^{2}}]\}$, which corresponds to a Gaussian
Ornstein-Uhlenbeck process \cite{Uhlenbeck823,Gillespie2084,Jing240403} and can be realized by pseudomode theory \cite{Barnett1997,Garraway2290,Garraway4636,Man062104,Man5763,Mazzola012104,Pleasance062105,Tamascelli030402}. Specially, in the wideband limit with $\lambda_{v}$ tending to infinity (i.e., $\lambda_{v} \rightarrow{\infty}$), the environments become memoryless. The spectral density approximately takes $J_{v}(\omega)\rightarrow{\Gamma_{v}/{2\pi}}$, or equivalently, $W(\omega)\rightarrow\sqrt{\Gamma_1/{2\pi}}$ and $G(\omega)\rightarrow\sqrt{\Gamma_2/{2\pi}}$. According to these, we have $f_{v}(t) = \Gamma_{v}\delta(t)$ and $\kappa_{v}(t)=i\sqrt{\Gamma_{v}}\delta(t)$. With environments initially prepared in vacuum state having $\langle K_v\rangle=\langle \hat K_v \rangle=0$, Eq.~(\ref{eqano}) becomes
\begin{eqnarray}
\dot{a}_s&=&i\Delta_s{a_s}+i g_s x_0({b+ {b}^*})a_s-i J_s a_2+\varepsilon_s e^{-i \Delta_1 t}-Y_1,\nonumber\\
\dot{a}_2&=&(i\Delta_l+i\Delta_{\rm{sag}}){a_2}-i J_s {a_s}-Y_2,\nonumber\\
\dot{b}&=&-(i\omega_b+\gamma_b)b+ i g_s x_0 {{a_s^*} }a_s,\label{eqano1B}\\
\dot{Y_1}&=&- {\lambda _1}{Y_1} + \frac{1}{2}{\Gamma _1}{\lambda _1}{a_s},\,\,\dot{Y_2}=-{\lambda _2}{Y_2} + \frac{1}{2}{\Gamma _2}{\lambda _2}{a_2},\nonumber
\end{eqnarray}
where $Y_1=\int_{0}^{t}d{\tau}a_s(\tau)f_{1}(t-\tau)$ and $Y_2=\int_{0}^{t}d{\tau}a_2(\tau)f_{2}(t-\tau)$.

\end{document}